# Measurement Error and Nonlinear Confounding: Structural Equation Modeling Versus Double Machine Learning

Eunjeong Song · Sehee Hong
Department of Education, Korea University, Seoul, Republic of Korea

Eunjeong Song, ORCID iD: https://orcid.org/0000-0002-2302-3227
Sehee Hong, ORCID iD: https://orcid.org/0000-0001-5468-8398

**Abstract**

A latent confounder measured by fallible indicators can be adjusted for with structural equation modeling (SEM), which corrects measurement error but requires a specified functional form, or with double machine learning (DML), which learns the form but ignores measurement error. We derive both population biases and the crossover reliability, the composite reliability above which DML with the indicators or their mean as inputs is less biased than linear SEM; it depends on the nonlinear share of confounding and falls when covariates predict the confounder. A 46-condition Monte Carlo study confirmed both biases and the predicted ordering. With gradient boosting, learning the conditional means raised the DML bias by up to 0.05 and moved the finite-sample crossover upward. A quadratic SEM was unbiased when its form was correct and removed two thirds of the linear-SEM bias under a saturating nonlinearity but little of it under a cubic one.

*Keywords:* structural equation modeling, double machine learning, measurement error, latent confounder, nonlinear confounding

A confounder is a common cause of treatment and outcome. In behavioral research it is often a construct measured by a few fallible indicators rather than observed directly. For example, prior motivation, measured by three questionnaire items, may affect both the hours a student spends in a tutoring program (the treatment) and later achievement (the outcome). Two problems arise. First, because the indicators contain measurement error, adjusting for them instead of for the construct leaves *residual confounding*, confounding that the adjustment does not remove (Cole & Preacher, 2014; Westfall & Yarkoni, 2016). Second, the construct can affect treatment and outcome nonlinearly, so a linear adjustment leaves residual confounding even without measurement error. This paper studies the two problems together for the confounder case; for a latent outcome see Gilbert (2025).

Structural equation modeling (SEM) addresses the first problem. A measurement model relates the indicators to the latent variable, and a structural model relates the latent variable to treatment and outcome. Because the measurement error is modeled, a correctly specified SEM estimates the structural coefficients consistently (Bollen, 1989). Nonlinear relations can be added through latent moderated structural equations (LMS; Klein & Moosbrugger, 2000), which estimate latent product terms by numerical integration. The analyst must, however, specify the functional form (the shape of the relation). If the true relation is nonlinear and a linear model is fitted, the structural coefficient converges to a *pseudo-true parameter*, the value that brings the misspecified model closest to the population distribution (White, 1982). Its distance from the true value does not shrink as the sample grows.

Double machine learning (DML) addresses the second problem. In the partially linear model $Y = \beta T + f(X) + e$, the treatment effect $\beta$ is estimated after *partialling out*, that is, after removing from $T$ and $Y$ their conditional means given the adjustment variables $X$ (Robinson, 1988). DML learns these means with flexible machine-learning methods, estimating them on one part of the sample and applying them to another (*cross-fitting*), and uses an *orthogonal* estimating equation that is insensitive to small errors in the learned means, so the usual standard errors remain valid (Chernozhukov et al., 2018). When the confounder is latent, the analyst can only supply *proxies* for it: the indicators, their mean, or a factor score; we call DML with such inputs *proxy-input DML*. However flexible its learner, any estimator that conditions on the proxies as if they were the confounder converges to the effect adjusted for the proxies. The measurement-error bias studied here is therefore a property of the adjustment set, not of DML. We use DML because it combines machine-learning predictions with valid standard errors for a low-dimensional effect.

Linear SEM and proxy-input DML each correct one source of bias and ignore the other; the practical question is which error is worse, and when. Each source has been documented separately: error-prone covariates bias regression and propensity-score adjustment (Steiner et al., 2011), and using sum scores, factor scores, or raw items as covariates leaves the estimate biased unless the measurement model is part of the estimation (Croon, 2002; Perrett & Kanopka, 2026; Rosseel & Loh, 2024). An omitted nonlinear term can leave the fit of a linear SEM unaffected (Mooijaart & Satorra, 2009). We know of no study that compares SEM and DML under one data-generating process with both sources present. Because the form of a nonlinear relation is rarely known in advance, the main comparison is between linear SEM and proxy-input DML; a quadratic SEM serves as a correctly specified benchmark, and we also ask how it performs when its form is wrong.

We make two contributions. First, we derive the population biases of linear SEM and proxy-input DML under linear and nonlinear confounding by a latent variable. The *crossover reliability*, at which proxy-input DML becomes less biased than linear SEM, depends on the share of confounding that is nonlinear and on the information that the proxies and covariates carry about the confounder. Second, a Monte Carlo study tests these predictions and measures what the formulas omit: the finite-sample cost of learning the conditional means and the bias of a misspecified quadratic SEM. The study answers three questions.

1. How do the bias, root mean squared error (RMSE), and coverage of linear SEM, quadratic SEM, and DML depend on reliability and on the form of confounding, and does the crossover reliability predict the finite-sample ordering?
2. How useful is a quadratic SEM when the true nonlinearity is not quadratic, and can the data choose between the linear and the quadratic model?
3. Do correlated covariates or unequal loadings change the ordering?

## Population Bias and the Crossover Reliability

### Model and Definitions

Figure 1 shows the data-generating model. Its structure (which variables affect which) is fixed; the design varies the functional form $g$ of the confounder's effect, the loadings, the correlation $\rho$ between the confounder and a covariate, and the sample size. A standardized latent confounder $U$ has three continuous indicators,

$$u_j = \lambda_j U + \epsilon_j, \qquad \epsilon_j \sim N(0, \theta_j), \quad j = 1,2,3. \qquad (1)$$

Ten observed covariates $\boldsymbol{Z}$ are multivariate normal with unit variances and correlations $.3^{|k-l|}$. The confounder $U \sim N(0,1)$ is independent of $\boldsymbol{Z}$ except in the correlated-covariate conditions, where $U = \rho Z_1 + \zeta_U$ with disturbance $\zeta_U \sim N(0, 1-\rho^2)$. A continuous treatment $T$ and outcome $Y$ follow

$$T = \gamma_T U + \kappa_T\, g(U) + \boldsymbol{b}_T' \boldsymbol{Z} + \zeta_T, \qquad (2)$$
$$Y = \beta T + \gamma_Y U + \kappa_Y\, g(U) + \boldsymbol{b}_Y' \boldsymbol{Z} + \zeta_Y, \qquad (3)$$

where $g$ is a centered function of $U$, uncorrelated with $U$, that represents the nonlinear part of confounding; any linear part of a nonlinear effect is absorbed in $\gamma_T$ and $\gamma_Y$. The estimand is $\beta$, the effect of a one-unit increase in $T$ on $Y$, which is the same for every unit; adjusting for $(U, \boldsymbol{Z})$ removes all confounding. Four forms are used: linear ($g = 0$), quadratic ($g = U^2 - 1$), a bounded *saturating* form ($g = U^2/(1 + U^2)$, centered), and an odd *cubic* form ($g = U^3 - 3U$). Under the last two forms a quadratic model is misspecified to different degrees: a quadratic term captures 69% of the variance of the saturating form and none of the cubic form. The design values are $\beta = 0.30$, $\gamma_T = \gamma_Y = 0.50$, and $\kappa_T = \kappa_Y = 0.30$ (quadratic), 1.603 (saturating), and 0.173 (cubic), so that the nonlinear part has variance 0.18 in every nonlinear form. The first three covariates affect $T$ and all ten affect $Y$ (coefficients 0.10). The indicator errors and the structural disturbances $\zeta_T, \zeta_Y \sim N(0,1)$, with variances $\psi_T = \psi_Y = 1$, are mutually independent and independent of $(U, \boldsymbol{Z})$, and the indicators have no direct effects on $T$ or $Y$. With these values $U$ accounts for 29% of the variance of $T$ in the nonlinear forms (17% linear, 12% nonlinear).

**Figure 1**

*The Data-Generating Model (Equations 1–3)*

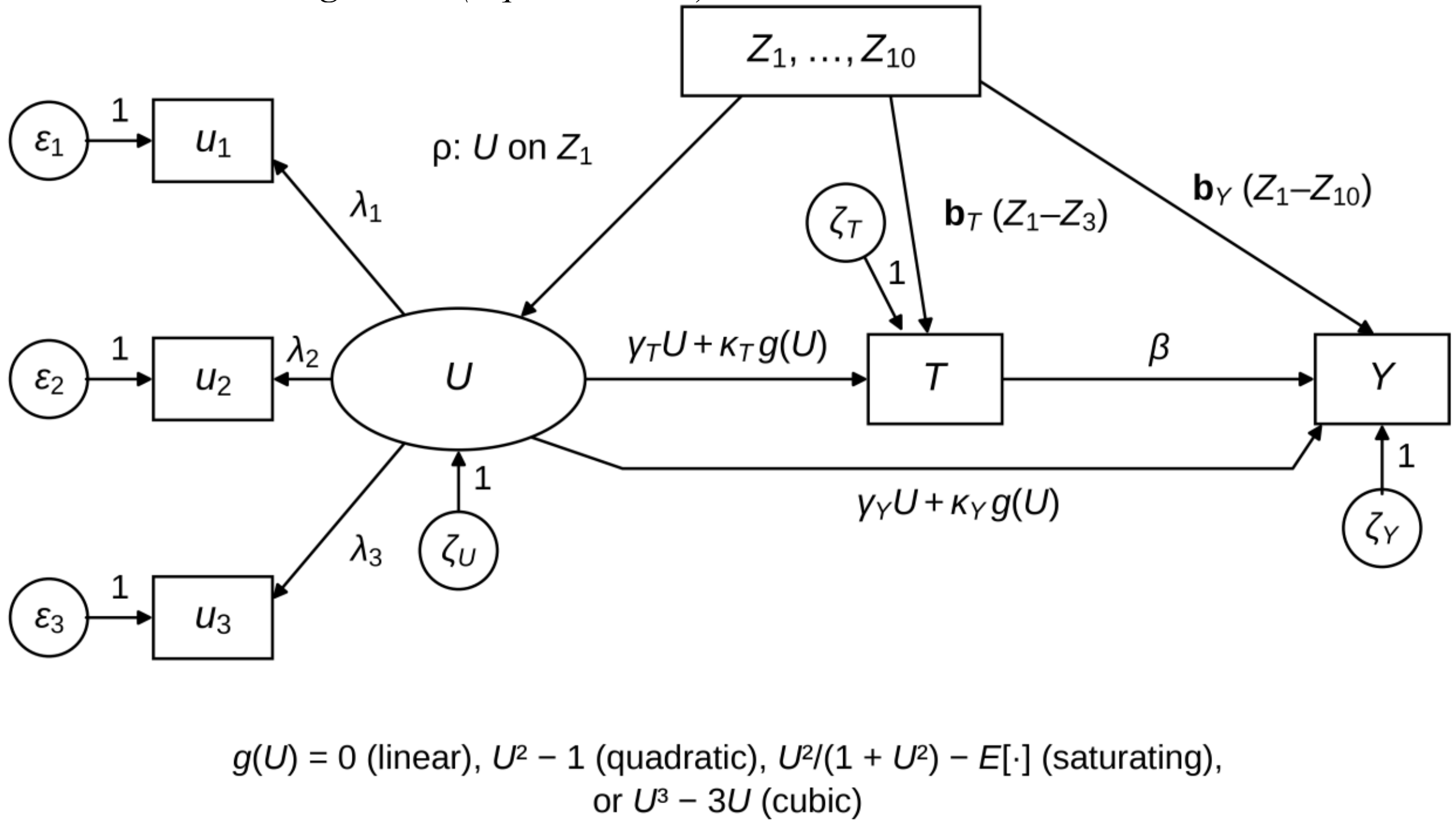

*Note.* Ellipse and circles: latent variable, indicator errors ($\epsilon$), and disturbances ($\zeta$); rectangles: observed variables; arrows: regressions; $\rho$ = corr($U$, $Z_1$), nonzero only in the correlated-covariate conditions, where Var($\zeta_U$) = $1 - \rho^2$.

Indicators are *parallel* when all loadings and all error variances are equal (Jöreskog, 1971); the indicators have unit variance, so $\theta_j = 1 - \lambda_j^2$. *Composite reliability* $\omega$ is the reliability of the indicator mean, $(\sum_j \lambda_j)^2/\{(\sum_j \lambda_j)^2 + \sum_j \theta_j\}$ (McDonald, 1999); with parallel indicators $\omega = 3\lambda^2/(1 + 2\lambda^2)$, so loadings of .6, .7, .8, and .9 give $\omega$ = .63, .74, .84, and .93. The *information* of the indicators about $U$ is $I = \sum_j \lambda_j^2/\theta_j$, the precision they add to the unit prior precision of $U$. The reliability of the optimally weighted composite, the *maximal reliability* $\omega_{\text{opt}} = I/(1 + I)$ (coefficient *H*; Hancock & Mueller, 2001), equals $\omega$ for parallel indicators and exceeds it otherwise.

**Population Bias of Each Approach**

The *population target* of an estimator is the value to which it converges as $N$ grows (for a misspecified model, the pseudo-true parameter); its *population bias* is that value minus $\beta$.

*Proxy-input DML.* Let $X$ be the adjustment set: the indicators or a score, plus the covariates. The population target of partialling-out is $\beta_X = E[(T - m)(Y - l)]/E[(T - m)^2]$ with $m = E(T \mid X)$ and $l = E(Y \mid X)$ (Robinson, 1988). Let $\tau^2$ be the variance of $U - E(U \mid X)$, the part of the confounder that the adjustment set does not capture, and let $V = \text{Var}[g(U) - E\{g(U) \mid X\}]$. Under joint normality of $(U, X)$ the cross term between $U - E(U \mid X)$ and $g(U)$ vanishes for any $g$ uncorrelated with $U$ (Supplement A.1), and

$$b_X = \beta_X - \beta = \frac{\gamma_T\gamma_Y\tau^2 + \kappa_T\kappa_Y V}{\psi_T + \gamma_T^2\tau^2 + \kappa_T^2 V}. \qquad (4)$$

Equation 4 is an omitted-variable bias (Chernozhukov et al., 2026) and holds when the conditional means are known; the additional bias from learning them is the *cost of learning*, measured in the simulation as the DML bias minus Equation 4. When $U$ is independent of $\boldsymbol{Z}$, $\tau^2$ is one minus the reliability of the best linear score in the adjustment set: $1 - \omega$ for the mean and $1 - \omega_{\text{opt}}$ for the indicators. The two coincide under parallel measurement, because the mean is then a sufficient statistic for $U$ (it carries all their information about $U$). For quadratic confounding with $\tau^2 = 1 - \omega$, $V = 2(1 - \omega^2)$; for the cubic form, $V = 6(1 - \omega^3)$ (Supplement A.3).

*Linear SEM (SEM-L).* Because $g(U)$ is uncorrelated with $U$, the indicators, and the covariates for any $\rho$ (Supplement A.4), a linear SEM reproduces the population covariance matrix exactly, with the treatment coefficient shifted to $\beta + b_L$ (the other coefficients of $Y$ shift accordingly),

$$b_L = \frac{\kappa_T\kappa_Y \mathrm{Var}(g)}{\psi_T + \kappa_T^2 \mathrm{Var}(g)}. \qquad (5)$$

The bias does not depend on reliability, because the measurement model is correct and the latent variable is controlled. Because the covariance matrix is reproduced exactly, chi-square tests and covariance-based fit indices cannot detect the omitted nonlinearity (Mooijaart & Satorra, 2009). With the design values, $b_L = 0.153$ for every nonlinear form (Table A1).

*Quadratic SEM (SEM-Q).* A latent model with $U$ and $U^2$ in both structural equations is correctly specified under linear and quadratic confounding, where its population bias is zero. Under the saturating and cubic forms it is misspecified; its population bias, approximated at $N$ = 20,000, is about 0.05 and 0.14–0.15 (Table C4).

**The Crossover Reliability and Two Extensions**

The crossover reliability $\omega^*$ is the reliability at which the two population biases are equal; below it, proxy-input DML is more biased than SEM-L, and above it, less. In the symmetric case $\gamma_T = \gamma_Y = \gamma$ and $\kappa_T = \kappa_Y = \kappa$, Equations 4 and 5 are equal, for any $\psi_T$, when $\gamma^2\tau^2 = \kappa^2\{\mathrm{Var}(g) - V\}$: the linear confounding that the adjustment set leaves equals the nonlinear confounding that it captures. Under quadratic confounding with $\rho = 0$ this gives

$$\omega^* = \frac{\sqrt{\gamma^4 + 8\kappa^2\gamma^2} - \gamma^2}{4\kappa^2}; \qquad (6)$$

for other forms, unequal coefficients, or $\rho \neq 0$, the equality is solved numerically (Supplement A.6). For the design values $\omega^* = .67$ under quadratic (Equation 6), $.71$ under saturating, and $.73$ under cubic confounding (Figure 2). In the symmetric case and for a given form and $\rho$, $\omega^*$ depends on the coefficients only through the *nonlinear share* $\pi = \kappa^2\mathrm{Var}(g)/\{\gamma^2 + \kappa^2\mathrm{Var}(g)\}$, the proportion of the variance of the confounder's effect, $\gamma U + \kappa\, g(U)$, that is nonlinear (Table 1; Figure 2C). Given $\omega$, parallel indicators carry the information $I = \omega/(1 - \omega)$ whatever their number, so their number does not change the population bias. The equality of Equations 4 and 5 also gives the crossover reliability for an interaction of $U$ with a covariate, $\omega^* = 1 - \pi$ (Supplement A.6).

Two extensions of Equation 4 motivate the correlated-covariate and unequal-loadings conditions. First, when $U$ is correlated with the covariates, the unexplained variance becomes

$$\tau^2 = \left\{ \frac{1}{1-\rho^2} + \sum_j \frac{\lambda_j^2}{\theta_j} \right\}^{-1}, \qquad (7)$$

the inverse of the total information about $U$: its prior precision (1), the indicator information $I$, and the covariate information $\rho^2/(1-\rho^2)$ (Supplement A.2). Because the equality of Equations 4 and 5 fixes $\tau^2$ for given coefficients, indicators and covariates substitute for each other (Sengewald & Pohl, 2019): correlations of .6 and .8 add 0.56 and 1.78 to the total information and move the quadratic crossover reliability from .67 to .60 and .22 (Table 1; Figure 2D). Second, under unequal loadings the mean is no longer sufficient: an optimally weighted score and a regression factor score with known measurement parameters share the $\tau^2 = 1 - \omega_{\text{opt}}$ of the indicators, whereas the mean has $\tau^2 = 1 - \omega$.

**Table 1**

*Crossover Reliability $\omega^*$ by the Nonlinear Share $\pi$ and corr($U$, $Z_1$)*

| $\pi$ | Quadratic | Saturating | Cubic | Quadratic, $\rho$ = .6 | Quadratic, $\rho$ = .8 |
|---|---|---|---|---|---|
| .10 | .91 | .91 | .91 | .90 | .89 |
| .20 | .83 | .84 | .85 | .81 | .75 |
| .30 | .76 | .78 | .79 | .72 | .57 |
| .42 | .67 | .71 | .73 | .60 | .22 |
| .50 | .62 | .66 | .68 | .51 | — |
| .67 | .50 | .55 | .59 | .30 | — |

*Note.* Symmetric case ($\gamma_T = \gamma_Y$, $\kappa_T = \kappa_Y$); $\pi$ = .42 is the design value. $\rho$ = 0 unless stated. SEM-L = linear SEM; proxy-input DML = DML with the indicators or their mean as inputs. — = proxy-input DML is less biased than SEM-L at every reliability.

**Figure 2**

*Population Bias and the Crossover Reliability*

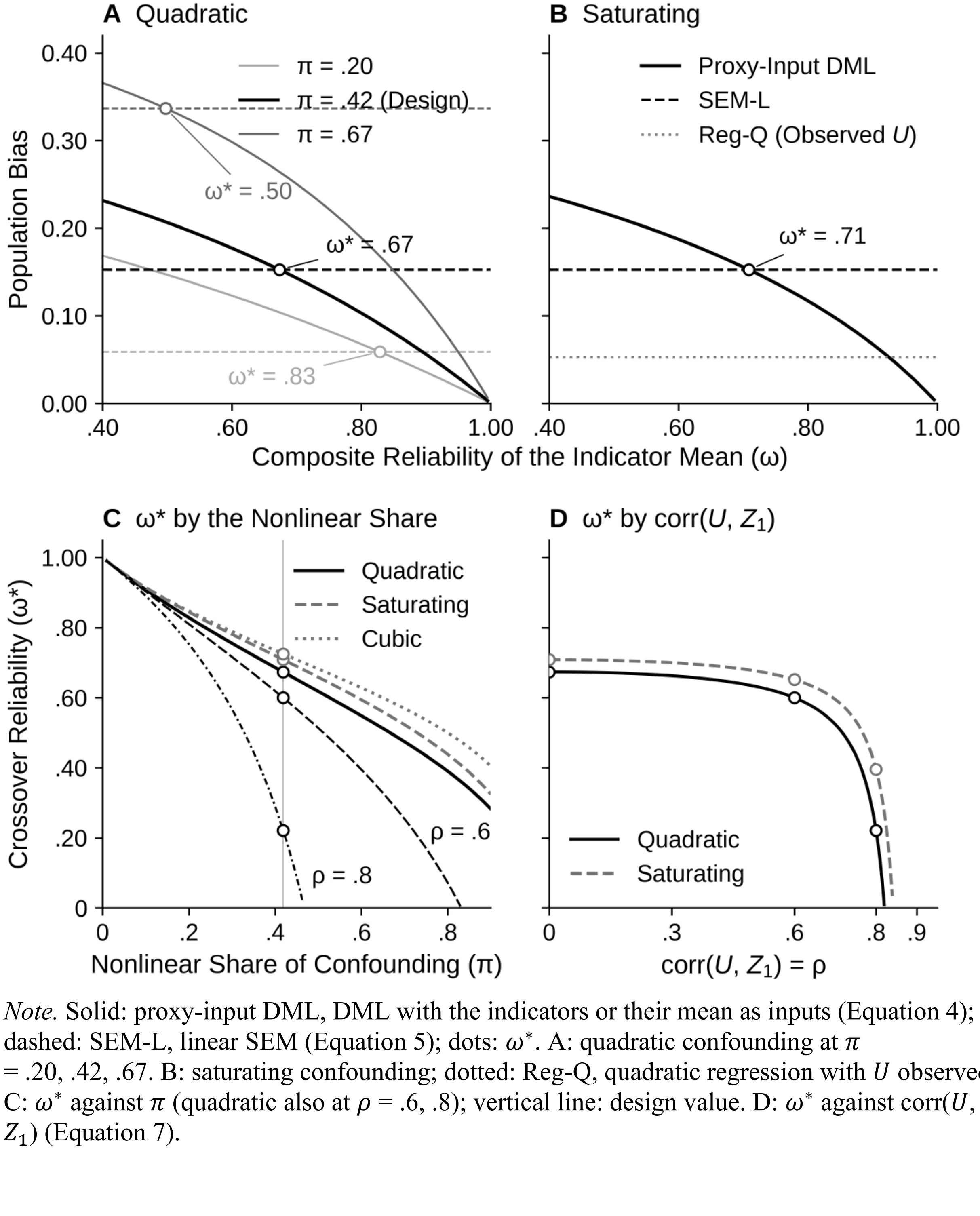

*Note.* Solid: proxy-input DML, DML with the indicators or their mean as inputs (Equation 4); dashed: SEM-L, linear SEM (Equation 5); dots: $\omega^*$. A: quadratic confounding at $\pi$ = .20, .42, .67. B: saturating confounding; dotted: Reg-Q, quadratic regression with $U$ observed. C: $\omega^*$ against $\pi$ (quadratic also at $\rho$ = .6, .8); vertical line: design value. D: $\omega^*$ against corr($U$, $Z_1$) (Equation 7).

# Method

## Conditions

Table 2 gives the 46 conditions, each with 500 replications (seeds and common random numbers: Supplement B.1). The core reliabilities bracket the crossover reliabilities (.67–.73; Table 1). Two conditions at $N = 1{,}000$ use the crossover reliability itself ($\omega$ = .67 and .71), where Equations 4 and 5 predict equal bias. In the unequal-loadings block the loadings are proportional to (0.5, 1, 1.4) at $\omega$ = .63 and to (0.8, 1, 1.15) at $\omega$ = .84, scaled so that $\omega$ equals its parallel counterpart; $\omega_{\text{opt}}$ = .76 and .90.

**Table 2**

*Simulation Design*

| Block | Factor | $N$ | Condition count |
|---|---|---|---|
| Core | linear, quadratic, saturating × $\omega$ = .63, .74, .84, .93 | 250, 1,000 | 24 |
| Crossover | quadratic at $\omega^*$ = .67; saturating at $\omega^*$ = .71 | 1,000 | 2 |
| Cubic form | cubic × $\omega$ = .63, .93 | 1,000 | 2 |
| Observed confounder | linear, quadratic, saturating; $U$ observed without error | 250, 1,000 | 6 |
| Correlated covariates | quadratic, saturating × $\omega$ = .63, .93 × $\rho$ = .6, .8 | 1,000 | 8 |
| Unequal loadings | quadratic, saturating × $\omega$ = .63, .84 | 1,000 | 4 |

*Note.* Indicators are parallel except in the unequal-loadings block. The core and crossover blocks form the *core conditions*; the other blocks are the *sensitivity conditions*. DML-FS and DML-Opt are added in the unequal-loadings block and its four matching core conditions.

## Estimators

*SEM-L* is a one-factor measurement model with the first loading fixed to 1 (Bollen, 1989) and linear regressions of $T$ and $Y$ on $U$ and the ten covariates, with $T$ also predicting $Y$. *SEM-Q* adds the latent quadratic term $U^2$ (XWITH in Mplus) to both structural equations and is estimated by LMS (30 integration points). In every latent-confounder condition $U$ is regressed on the ten covariates, so that $U$ can covary with $\boldsymbol{Z}$. Both models use the MLR estimator of Mplus (maximum likelihood with robust standard errors; Satorra & Bentler, 1994; Yuan & Bentler, 2000), whose sandwich standard errors allow for the nonnormality induced by the omitted nonlinear term. SEM-L is nested in SEM-Q; the two models, specified in advance, are compared by the *scaled difference test* (the likelihood-ratio test scaled for MLR; Satorra & Bentler, 2001), whose *difference scaling factor* $c_d$ combines the two models' scaling corrections (Supplement B.2). *SEM-Select* reports SEM-Q when this test rejects at $\alpha = .05$ and SEM-L otherwise, including when SEM-Q cannot be fitted or when $c_d \leq 0$ leaves the test undefined (Satorra & Bentler, 2010).

*DML-Items* uses the partially linear partialling-out score with the three indicators and the ten covariates as adjustment variables; *DML-Mean* replaces the indicators by their mean. The conditional means of $T$ and $Y$ are learned by histogram-based gradient boosting (Friedman, 2001) in scikit-learn (Pedregosa et al., 2011), with five-fold cross-fitting in DoubleML (Bach et al., 2022). For each condition and estimator, hyperparameters were selected once, from 24 candidates of the grid in Table B1, by three-fold cross-validated prediction error on five separate tuning datasets (Bach et al., 2024), and then held fixed for the 500 replications (Supplement B.3). *DML-FS* uses the regression factor score (the expected value of $U$ given the indicators; Grice, 2001) of a one-factor model estimated within each training fold, and *DML-Opt* the score with the population weights $\lambda_j/\theta_j$, unknown in practice. Both scores enter DML as observed covariates and therefore keep the bias of Equation 4.

In the observed-confounder block, *Reg-L* and *Reg-Q* are ordinary least squares with $U$, or $U$ and $U^2$, and the covariates, and *DML-True* is DML with the observed $U$. Reg-L isolates the cost of linearity (its bias follows Equation 5) and DML-True the cost of learning.

## Evaluation Criteria

For converged, proper solutions (no negative variances or nonpositive-definite matrices), we report the bias, the relative bias (acceptable within ±5%; Hoogland & Boomsma, 1998), the RMSE, and the *coverage*, the proportion of 95% intervals containing $\beta$. Mean standard errors within ±10% of the empirical standard deviation count as accurate. The Monte Carlo standard error (MCSE; Morris et al., 2019) of a coverage of .95 is .010. Differences between estimators are paired and interpreted only beyond two Monte Carlo standard errors; the ordering of DML and SEM-L is called predicted only when the paired difference has the predicted sign and exceeds that margin. Failed fits are counted, not repaired (Supplement B.5).

## Software and Preregistration

SEM-L and SEM-Q were estimated with Mplus 9.1 (Muthén & Muthén, 2026) and DML with DoubleML 0.11.4 and scikit-learn 1.8.0 in Python 3.13. The study was preregistered on 24 February 2026; the protocol was amended on 10 October 2026 and deposited on the Open Science Framework (OSF) before the revised analyses (anonymized view-only link: https://osf.io/kwnfd/overview?view_only=fc7375b1f4074712b5652a1b15db2d0e); Supplement D summarizes the amendment.

# Results

## Core Conditions

Table 3 gives the bias and Figure 3 the bias and coverage in the core conditions, each with the population bias of Equation 4; RMSE followed the same ordering as bias (Figure C1). Every fit in the core conditions converged to a proper solution, and the mean standard errors were within 10% of the empirical standard deviations for every estimator except SEM-Select (Supplement C), so coverage below .95 reflects bias. Proxy-input DML never met the 5% criterion for relative bias. Below, DML without a suffix denotes DML-Items and DML-Mean together.

Under linear confounding, SEM-L and SEM-Q were unbiased (|bias| ≤ 0.007) with coverage .95–.97. DML-Items and DML-Mean were biased as Equation 4 predicts, exceeding it by 0.001–0.008 at $N$ = 1,000.

Under quadratic confounding the ordering predicted by $\omega^*$ held at $N$ = 1,000. The bias of SEM-L was 0.147–0.153 at every reliability and sample size (Equation 5: 0.153). Its coverage matched the value implied by its bias and standard error (Figure 3; Supplement A.7), and its fit indices were indistinguishable from those under linear confounding (Table C2). SEM-Q was unbiased (|bias| ≤ 0.007) with coverage .94–.96. The DML bias exceeded Equation 4 by 0.003–0.013 at $N$ = 1,000 and by 0.025–0.046 at $N$ = 250. At $N$ = 1,000 DML-Mean was more biased than SEM-L at $\omega = .63$ (paired difference 0.020, MCSE 0.001) and less biased at $\omega = .74$ (0.019), as $\omega^* = .67$ predicts. At $\omega^*$ itself, where the population biases are equal, DML-Mean was more biased by 0.005 (MCSE 0.001) and DML-Items by 0.009. The *finite-sample crossover*, the interval between simulated reliabilities in which the observed ordering changed, therefore lay between $\omega = .67$ and .74 at $N = 1{,}000$; at $N = 250$ the larger cost moved it to between .74 and .84.

Under saturating confounding, SEM-L again had a bias of 0.148–0.154. SEM-Q, now misspecified, had a bias of 0.043–0.051, one third of SEM-L's, with coverage .64–.72 at $N = 1{,}000$. DML-Mean was more biased than SEM-L at $\omega^* = .71$ (by 0.002, MCSE 0.0007) and less biased at $\omega = .74$ (by 0.009). The finite-sample crossover therefore lay between .71 and .74 at $N = 1{,}000$ and between .74 and .84 at $N = 250$; DML remained more biased than SEM-Q at every reliability.

With the confounder observed, DML-True had a bias of 0.009–0.017 at $N = 1{,}000$ and 0.031–0.043 at $N = 250$ across the three forms, the cost of learning alone (Table C6a). The cost of learning was positive in all 102 DML results (95 beyond two Monte Carlo standard errors), consistent with the regularization bias of an underfitted learner (Chernozhukov et al., 2018), which leaves part of the confounder's signal in both residuals, so the cost takes the sign of the confounding. In the core conditions, DML-Items and DML-Mean differed by at most 0.007 at $N = 1{,}000$ and 0.022 at $N = 250$ (Figure 3).

**Table 3**

*Bias by Composite Reliability in the Core Conditions (N = 1,000)*

| Form | Estimator | ω = .63 | ω* (.67/.71) | ω = .74 | ω = .84 | ω = .93 |
|---|---|---|---|---|---|---|
| Linear | SEM-L | 0.000 | — | 0.001 | 0.002 | 0.002 |
| | SEM-Q | 0.000 | — | 0.001 | 0.002 | 0.002 |
| | DML-Items | 0.092 | — | 0.067 | 0.041 | 0.025 |
| | DML-Mean | 0.086 | — | 0.063 | 0.041 | 0.026 |
| | Equation 4 | 0.085 | — | 0.060 | 0.038 | 0.018 |
| Quadratic | SEM-L | 0.152 | 0.152 | 0.152 | 0.153 | 0.153 |
| | SEM-Q | 0.001 | 0.000 | 0.001 | 0.001 | 0.002 |
| | DML-Items | 0.176 | 0.161 | 0.136 | 0.095 | 0.055 |
| | DML-Mean | 0.171 | 0.157 | 0.134 | 0.094 | 0.054 |
| | Equation 4 | 0.168 | 0.153 | 0.127 | 0.084 | 0.041 |
| Saturating | SEM-L | 0.153 | 0.153 | 0.153 | 0.154 | 0.154 |
| | SEM-Q | 0.051 | 0.051 | 0.051 | 0.050 | 0.051 |
| | DML-Items | 0.182 | 0.157 | 0.150 | 0.109 | 0.064 |
| | DML-Mean | 0.181 | 0.155 | 0.144 | 0.103 | 0.059 |
| | Equation 4 | 0.179 | 0.153 | 0.140 | 0.098 | 0.051 |

*Note.* Bias = mean estimate − β (β = 0.30); Monte Carlo standard errors ≤ 0.002. SEM-L and SEM-Q = linear and quadratic SEM; DML-Items and DML-Mean = DML with the three

indicators or their mean as inputs. Equation 4 = population bias of DML-Items and DML-Mean; that of SEM-L is 0 (linear) or 0.153 (Equation 5). ω* = crossover reliability (not run for linear confounding). Text differences use unrounded values.

**Figure 3**

*Bias and Coverage in the Core Conditions*

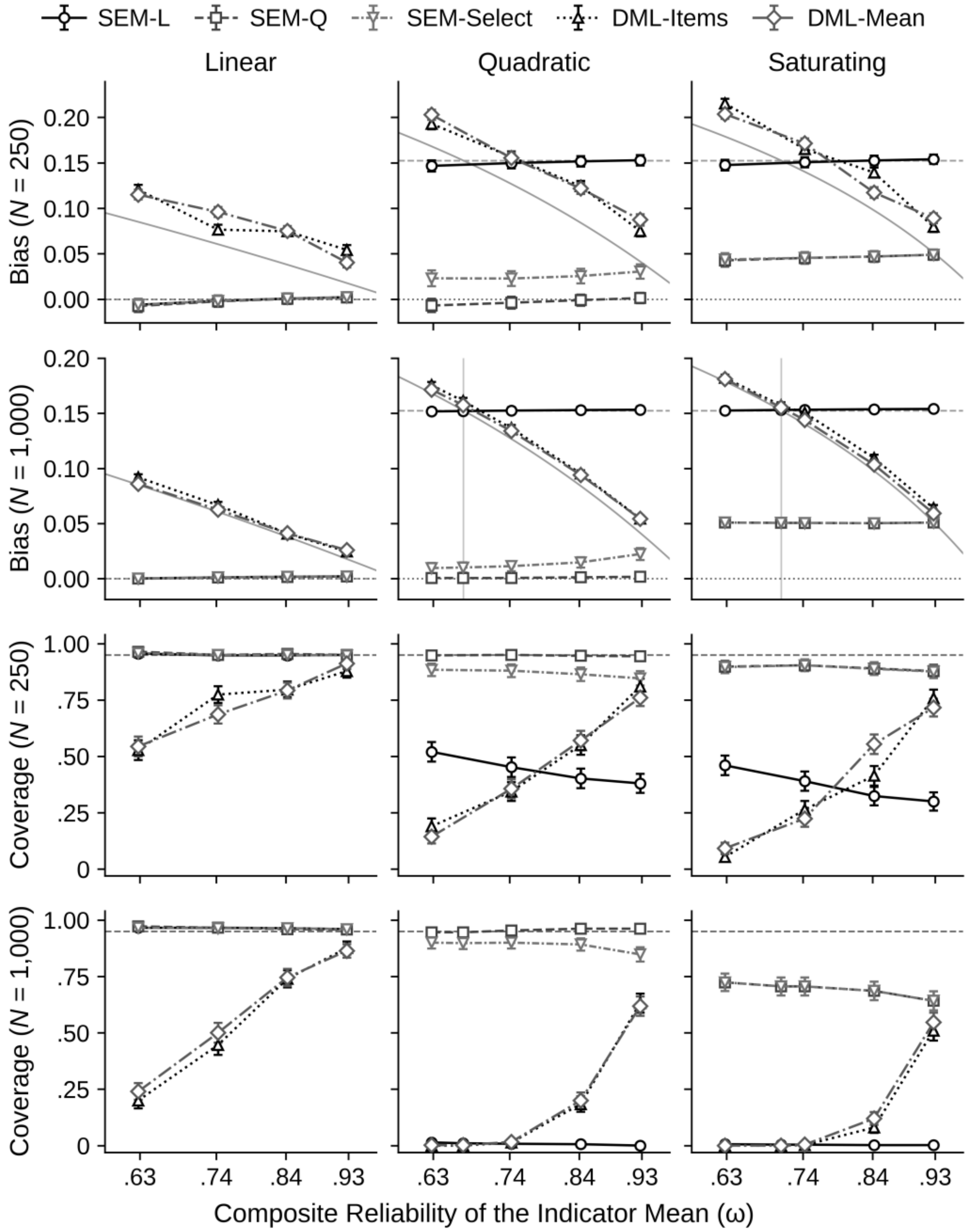

*Note.* SEM-L and SEM-Q = linear and quadratic SEM; SEM-Select = SEM-Q when the scaled difference test rejects, otherwise SEM-L; DML-Items and DML-Mean = DML with the three indicators or their mean as inputs. Error bars: 1.96 Monte Carlo standard errors; grey lines: population biases (Figure 2); vertical line: $\omega^*$, the crossover reliability; dashed line in the coverage rows: .95.

**Misspecified Nonlinear Forms and Model Selection**

Under the cubic form, SEM-Q had a bias of 0.128 at $\omega$ = .63 and 0.145 at $\omega$ = .93, only 0.022 and 0.007 below SEM-L's (Table C6d). The cost of learning was largest under this form at $N$ = 1,000: DML exceeded Equation 4 by 0.012–0.014 at $\omega$ = .63 and by 0.038–0.045 at $\omega$ = .93, where the conditional mean is a steep cubic function that shallow trees approximate poorly. DML nevertheless became less biased than SEM-L between the two reliabilities, as $\omega^* = .73$ predicts.

The scaled difference test detected the quadratic and saturating forms but rarely the cubic form, where $c_d$ was about 4–5 (Supplement B.2). At $N$ = 1,000 SEM-Select chose SEM-Q in 5–6% of the linear, 88–95% of the quadratic, 100% of the saturating, and 29% and 14% of the cubic replications (Table C3). Under quadratic confounding the test rejected in more than 99% of the replications in which it was defined but was undefined ($c_d \leq 0$) in 5–15% of them (Supplement B.2); SEM-Select then reported SEM-L, which accounts for its bias of 0.010–0.030 under this form. Because SEM-Select mixes estimates from two models, its mean standard error fell 17–48% short of its empirical standard deviation and its coverage was .85–.91 (Tables C1a, C1b, and C6).

**Correlated Covariates and Unequal Loadings**

Covariate information and unequal loadings changed the DML bias as Equations 4 and 7 predict and left the SEM biases unchanged (Figure 4; Table C6). When $U$ was correlated with $Z_1$, the DML-Mean bias at $\omega$ = .63 fell from 0.171 ($\rho$ = 0) to 0.152 ($\rho$ = .6) and 0.122 ($\rho$ = .8) under quadratic confounding, against 0.168, 0.146, and 0.113 from Equations 4 and 7. At $\rho$ = .8 DML was less biased than SEM-L at both reliabilities, as the crossover reliabilities of .22 (quadratic) and .40 (saturating; Supplement A.6) predict. At $\rho$ = .6 and $\omega$ = .63 (quadratic), the cost of learning absorbed the predicted gap of 0.007 in favor of DML: DML-Mean did not differ from SEM-L beyond Monte Carlo error, and DML-Items was more biased by 0.003. Under parallel loadings the mean, the factor score, and the optimal score gave biases within 0.002 of one another. Under unequal loadings matched on $\omega$, DML-Items, DML-FS, and DML-Opt were less biased than DML-Mean by 0.040–0.046 at $\omega$ = .63 and by 0.024–0.027 at $\omega$ = .84, within 0.01 of the differences that Equation 4 predicts (Tables A1 and C6c); the more informative inputs had a larger cost of learning. Because the indicators then have $\omega_{\text{opt}}$ = .76 at $\omega$ = .63, DML-Items was less biased than SEM-L there although $\omega < \omega^*$. DML-FS failed in 7% of the replications at $\omega$ = .63 because the one-factor model of a training fold did not converge (a Heywood case; Supplement B.3).

**Figure 4**

*Correlated Covariates and Unequal Loadings (N = 1,000)*

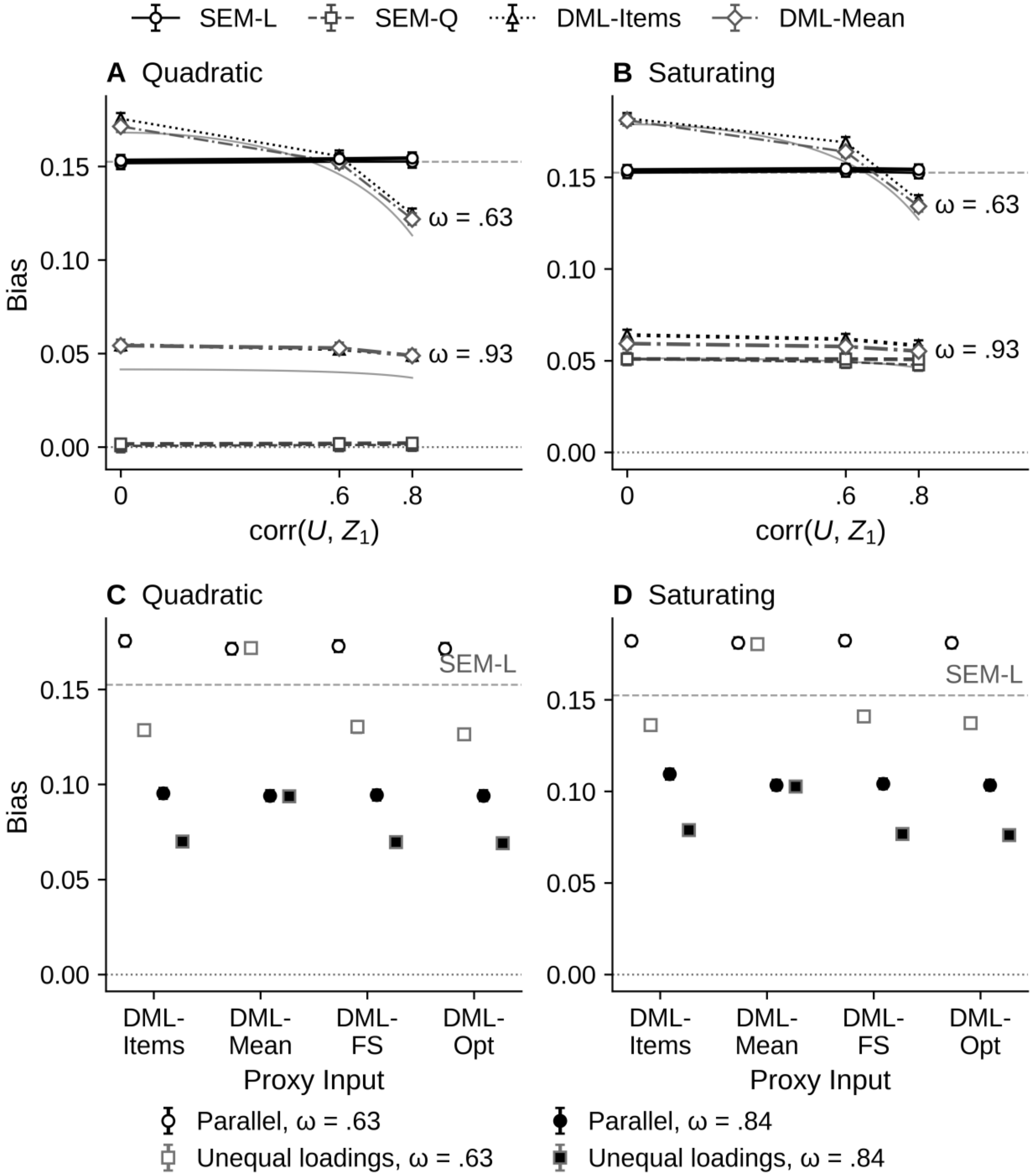

*Note.* SEM-L and SEM-Q = linear and quadratic SEM; DML-Items, DML-Mean, DML-FS, and DML-Opt = DML with the three indicators, their mean, regression factor scores, or the optimally weighted score as inputs. A and B: bias against corr($U$, $Z_1$) at $\omega$ = .63 and .93; grey lines: population biases (Equations 4, 5, and 7). C and D: proxy inputs under parallel and unequal loadings at matched $\omega$; dashed line: population bias of SEM-L (Equation 5).

# Discussion

## Summary

Three results answer the research questions. (1) The crossover reliability predicted the ordering of SEM-L and proxy-input DML at $N = 1{,}000$; with the learner used here, the cost of learning moved the finite-sample crossover upward, more at $N = 250$. (2) SEM-Q recovered the effect when its form was right, removed two thirds of the SEM-L bias under the saturating form, and little of it under the cubic form, which the scaled difference test rarely detected. (3) Covariates that predict the confounder lowered the DML bias as Equations 4 and 7 predict and left the SEM biases unchanged; under unequal loadings the reliability that governs DML on the indicators is $\omega_{\text{opt}}$, not $\omega$.

## Implications for Practice

In the symmetric case, two quantities decide which of SEM-L and proxy-input DML has the smaller population bias: the information about the confounder and the nonlinear share $\pi$. After the measurement model has been checked (Anderson & Gerbing, 1988), SEM-L estimates both information quantities: $\omega$ ($\omega_{\text{opt}}$ for the indicators under unequal loadings) and the $R^2$ of the latent regression of $U$ on $\boldsymbol{Z}$, which replaces $\rho^2$ in Equation 7. Both are consistent even when the nonlinear term is omitted (Supplement A.4). Because $\pi$ is rarely known, the comparison is a sensitivity analysis. The analyst posits a range of $\pi$ (e.g., .1 to .5) and reads $\omega^*$ for that range from Table 1 or computes it as in Supplement A.6, which also covers unequal coefficients, with their ratios as further sensitivity parameters. Proxy-input DML is less biased where $\omega$ exceeds every $\omega^*$ in the range, SEM-L where it falls below every $\omega^*$; in between, report both. Suppose that in the tutoring example the three motivation items have $\omega = .85$: SEM-L is favored as long as $\pi$ stays below .17 (quadratic) to .20 (cubic), and covariates that explain 36% of the variance of motivation lower these values only to .16–.18.

When an even (U-shaped) nonlinearity is plausible, SEM-Q is a third option: it was less biased than proxy-input DML in every condition except the cubic form at $\omega = .93$, and its estimates of $\gamma$ and $\kappa$ give the quadratic part of $\pi$ directly, $\pi = 2\kappa^2/(\gamma^2 + 2\kappa^2)$ with $U$ standardized, but it cannot capture an odd nonlinearity. The choice between SEM-Q and SEM-L should not rest on the scaled difference test alone, which is unstable in both directions when the two models differ by few parameters relative to their size (Supplement B.2); alternatives, not evaluated here, are a Wald test of the same restriction (the two quadratic coefficients) and the strictly positive statistic of Satorra and Bentler (2010).

## Limitations

The derivations hold for the model of Equations 1–3 under joint normality, which makes $E(U \mid X)$ linear and the cross terms of Equation 4 vanish; for a nonnormal construct they need not. The simulation results hold for the conditions examined: a single normally distributed latent confounder measured by three continuous indicators with independent errors and no direct effects on $T$ or $Y$; nonlinear confounding through $U$ alone; and a homogeneous effect of a continuous treatment. The simulation used one strength of nonlinear confounding and one learner, tuned on independent datasets from the same condition; cross-validation targets prediction error rather than the bias of the orthogonal score (Bach et al., 2024), and a wider grid did not reduce the cost of learning (Supplement B.3). Another learner would change this cost but not Equation 4, the limit of any learner whose predictions converge to the conditional means.

With heterogeneous effects, partialling-out estimates an average effect weighted by $\mathrm{Var}(T \mid X)$. The interval of SEM-Select does not account for the selection step. Open questions are whether factor-score corrections (Croon, 2002; Rosseel & Loh, 2024) can be combined with a machine-learning adjustment and how proximal inference (Miao et al., 2018) compares.

## Statements and Declarations

### Data and Code Availability

Code and summary results are available in Harvard Dataverse (https://doi.org/10.7910/DVN/IAUEMG).

### Funding

No funding was received for this study.

### Competing Interests

The authors declare no competing interests.

### Acknowledgments

DeepL was used for translation and English editing and OpenAI Codex to check the analysis code; the authors reviewed all outputs and take full responsibility for them.

### Ethics

The study uses simulated data only; no ethical review was required.

## Supplementary Materials

Measurement Error and Nonlinear Confounding: Structural Equation Modeling Versus Double Machine Learning

Supplement A gives the derivations, Supplement B the estimation protocol, Supplement C the result tables, and Supplement D the preregistration. Notation, equation numbers, and figure numbers refer to the main text; the analysis package provides code and summary results for reproducing the tables and figures.

## Supplement A. Derivations

### A.1 Population Bias of Partialling-Out With a Proxy Adjustment Set (Equation 4)

Let $X$ be the adjustment set (indicators or a score, plus $\boldsymbol{Z}$), $m(X) = E(T \mid X)$ and $l(X) = E(Y \mid X)$. Write $d = U - E(U \mid X)$ and $h = g(U) - E\{g(U) \mid X\}$. From Equations 2 and 3,

$$T - m = \gamma_T d + \kappa_T h + \zeta_T, \qquad (Y - l) - \beta(T - m) = \gamma_Y d + \kappa_Y h + \zeta_Y,$$

so the partialling-out target $\beta_X = E[(T - m)(Y - l)]/E[(T - m)^2]$ equals $\beta$ plus

$$\frac{\gamma_T\gamma_Y E(d^2) + (\gamma_T\kappa_Y + \gamma_Y\kappa_T)E(dh) + \kappa_T\kappa_Y E(h^2)}{\gamma_T^2 E(d^2) + 2\gamma_T\kappa_T E(dh) + \kappa_T^2 E(h^2) + \psi_T}.$$

Because $U$ and the variables in $X$ are jointly normal, $\mu = E(U \mid X)$ is linear in $X$ and $d = U - \mu$ is independent of $X$. Since $\mathrm{Cov}(d, U) = \mathrm{Var}(U) - \mathrm{Var}(\mu) = \tau^2$, $d = \tau^2 U + w$ with $w$ independent of $U$, and $E\{d\, g(U)\} = \tau^2 E\{Ug(U)\} + E(w)E\{g(U)\} = 0$ because $g$ is centered and uncorrelated with $U$. As $d$ is independent of every function of $X$, $E(dh) = E\{dg(U)\} = 0$: the cross terms vanish, and Equation 4 follows with $\tau^2 = E(d^2)$ and $V = E(h^2)$. Nothing in the argument uses the form of $g$: any nonlinear function of $U$ is its linear projection, absorbed in $\gamma_T$ and $\gamma_Y$, plus a part uncorrelated with $U$, to which Equation 4 applies.

### A.2 Information About *U* in the Indicators, the Mean, and the Covariates (Equation 7)

With $U = \rho Z_1 + \zeta_U$ and $\zeta_U \sim N(0, 1 - \rho^2)$, $U \mid \boldsymbol{Z} \sim N(\rho Z_1, 1 - \rho^2)$. Combining this prior with the indicator likelihood gives

$$\tau^2_{\text{items}} = \left\{\frac{1}{1 - \rho^2} + \sum_j \frac{\lambda_j^2}{\theta_j}\right\}^{-1}, \qquad E(U \mid \boldsymbol{u}, \boldsymbol{Z}) = \tau^2_{\text{items}}\left\{\frac{\rho Z_1}{1 - \rho^2} + \sum_j \frac{\lambda_j u_j}{\theta_j}\right\}:$$

the total precision is the prior precision of $U$ (1) plus the indicator information $I = \sum_j \lambda_j^2/\theta_j$ plus the covariate information $\rho^2/(1 - \rho^2)$. With several covariates, $1/(1 - \rho^2)$ is replaced by $1/(1 - R^2)$, where $R^2$ is the squared multiple correlation of $U$ with $\boldsymbol{Z}$ (in SEM-L, the $R^2$ of the latent regression of $U$ on $\boldsymbol{Z}$). For the mean $W = J^{-1}\sum_j u_j$, $I$ is replaced by $(\sum_j \lambda_j)^2/\sum_j \theta_j$. With $\rho = 0$ this gives $\tau^2 = 1 - \omega$ for the mean under any loading pattern and $\tau^2 = 1 - \omega_{\text{opt}}$ for the indicators, where $\omega_{\text{opt}} = I/(1 + I)$ is the maximal reliability (Hancock & Mueller, 2001). By the Cauchy–Schwarz inequality $I \geq (\sum_j \lambda_j)^2/\sum_j \theta_j$, with equality if and only if $\lambda_j/\theta_j$ is constant across indicators, as it is for parallel indicators: the mean is then a sufficient statistic for $U$. Otherwise the optimal linear score weights indicator $j$ by $\lambda_j/\theta_j$, and the regression factor score, which uses the same weights up to a constant, has the same $\tau^2$ when the measurement parameters are known.

**A.3 Residual Nonlinear Variance $V$**

Because $\mathrm{Var}(U) = 1$, $\mathrm{Var}(\mu) = 1 - \tau^2$. For $g(U) = U^2 - 1$, $E\{g(U) \mid X\} = \mu^2 + \tau^2 - 1$ and $V = 2\{1 - (1 - \tau^2)^2\}$, which is $2(1 - \omega^2)$ when $\tau^2 = 1 - \omega$. For $g(U) = U^3 - 3U$, $E\{g(U) \mid X\} = \mu^3 - 3(1 - \tau^2)\mu$ has variance $6(1 - \tau^2)^3$, so $V = 6\{1 - (1 - \tau^2)^3\} = 6(1 - \omega^3)$; $\mathrm{Var}(g) = 6$ and $\kappa_T = \kappa_Y = \sqrt{0.18/6} = 0.17321$. For the saturating form $g(U) = U^2/(1 + U^2) - E\{U^2/(1 + U^2)\}$, $E\{g(U) \mid X\}$ is evaluated by 160-node Gauss–Hermite quadrature and $V$ follows from the law of total variance; the marginal moments are $E\{U^2/(1 + U^2)\} = 0.34432$, $\mathrm{Var}(g) = 0.070084$, and $E\{g(U)(U^2 - 1)\} = 0.31136$, and $\kappa_T = \kappa_Y = \sqrt{0.18/\mathrm{Var}(g)} = 1.60260$ gives the nonlinear part the variance 0.18 of the other forms.

**A.4 Linear SEM Reproduces the Population Covariance Matrix (Equation 5)**

Because $g(U)$ is centered and uncorrelated with $U$, and each covariate can be written as $Z_k = r_k U + W_k$ with $W_k$ independent of $U$, $E\{Z_k g(U)\} = r_k E\{U g(U)\} = 0$; likewise $E\{u_j g(U)\} = \lambda_j E\{U g(U)\} = 0$, and $g(U)$ is independent of the indicator and structural errors. This holds whether or not $\rho = 0$ and for every form used (for the even forms these are the vanishing odd moments of a symmetric distribution; for the cubic form $E\{U(U^3 - 3U)\} = 3 - 3 = 0$). Writing $v = \mathrm{Var}(g)$ and $b_L = \kappa_T \kappa_Y v/(\psi_T + \kappa_T^2 v)$, the linear model with

$$\beta^* = \beta + b_L, \quad \gamma_Y^* = \gamma_Y - \gamma_T b_L, \quad \boldsymbol{b}_Y^* = \boldsymbol{b}_Y - b_L \boldsymbol{b}_T, \quad \mathrm{Var}(\zeta_T^*) = \psi_T + \kappa_T^2 v, \quad \mathrm{Var}(\zeta_Y^*) = \psi_Y + b_L^2 \psi_T + (\kappa_Y - b_L \kappa_T)^2 v,$$

and unchanged measurement parameters, treatment paths, and latent regression on $\boldsymbol{Z}$ reproduces every element of the population covariance matrix of $(u_1, u_2, u_3, \boldsymbol{Z}, T, Y)$. Hence the pseudo-true treatment coefficient of SEM-L is $\beta + b_L$ regardless of reliability and of $\rho$, the population fit of the linear model is perfect, and its measurement parameters and latent regression on $\boldsymbol{Z}$, and with them $\omega$ and $R^2$, are consistent. The same expression is the bias of Reg-L.

**A.5 Reg-Q Under the Saturating and Cubic Forms**

For observed $U$, the projection of $g$ on $U^2 - 1$ has coefficient $E\{g(U)(U^2 - 1)\}/2$. For the saturating form the residual variance is $v_Q = \mathrm{Var}(g) - [E\{g(U)(U^2 - 1)\}]^2/2 = 0.02161$, 30.8% of $\mathrm{Var}(g)$, and the Reg-Q bias is $\kappa_T \kappa_Y v_Q/(\psi_T + \kappa_T^2 v_Q) = 0.0526$. For the cubic form $E\{g(U)(U^2 - 1)\} = 0$, so Reg-Q keeps the full bias of Equation 5. These are population biases of observed-variable least-squares projections; the pseudo-true parameter of a latent quadratic model under a misspecified form is not the same projection and is approximated by fits at $N$ = 20,000 (Table C4).

**A.6 The Crossover Reliability (Equation 6)**

With $\gamma_T = \gamma_Y = \gamma$, $\kappa_T = \kappa_Y = \kappa$, quadratic confounding and $\rho = 0$, Equation 4 equals Equation 5 when $\gamma^2(1 - \omega) = \kappa^2\{\mathrm{Var}(g) - V\} = 2\kappa^2\omega^2$; $\psi_T$ cancels, and solving $2\kappa^2\omega^2 + \gamma^2\omega - \gamma^2 = 0$ for $\omega \in (0,1)$ gives Equation 6. For general coefficients (where $\psi_T$ remains in the equation), the saturating or cubic form, or $\rho \neq 0$ (with $\tau^2$ from Equation 7), the equality of Equations 4 and 5 is solved numerically for $\omega \in (0,1)$ by a bracketing root search. For the design coefficients this gives $\omega^*$ = .673 (quadratic), .709 (saturating), and .725 (cubic), and, against $\rho$ (Figure 2D), .673, .663, .600, .524, and .222 at $\rho$ = 0, .3, .6, .7, and .8 under quadratic confounding and .709, .700, .652, .595, and .396 under the saturating form; at $\rho$ = .9 the proxy-input bias is below the SEM-L bias at every reliability. Because $V$ depends on the adjustment set only through $\tau^2$, the equality is a condition on $\tau^2$ alone: the crossover corresponds to a fixed unexplained variance, $\tau^{*2}$ = .327, .291, and .275 for the three forms at the design values, and hence to a fixed total information $1/\tau^{*2} - 1$ (2.06, 2.43, and 2.64), of which the covariate contributes $\rho^2/(1-\rho^2)$ and the indicators the rest; at $\rho$ = .6 under quadratic confounding, $I$ = 2.06 − 0.56 = 1.50 and $\omega^*$ = .600.

*The nonlinear share.* Dividing the symmetric-case condition $\gamma^2\tau^2 = \kappa^2\{\mathrm{Var}(g) - V(\tau^2)\}$ by $\gamma^2$ and writing $a = \pi/(1-\pi) = \kappa^2\mathrm{Var}(g)/\gamma^2$ gives $\tau^2 = a\{1 - V(\tau^2)/\mathrm{Var}(g)\}$. The ratio $V/\mathrm{Var}(g)$ depends only on $\tau^2$ for a given form, and $\tau^2$ only on $\omega$ and $\rho$ (Equation 7 with $I = \omega/(1-\omega)$ for parallel indicators), so $\omega^*$ is a function of $(\pi, \rho)$ alone for each form; Table 1 evaluates it numerically, and under quadratic confounding with $\rho = 0$ it is $\omega^* = \{(1 + 4a)^{1/2} - 1\}/(2a)$, which is Equation 6 with $a = 2\kappa^2/\gamma^2$. The design value is $\pi$ = 0.18/(0.25 + 0.18) = .42. For asymmetric coefficients ($\gamma_T \neq \gamma_Y$ or $\kappa_T \neq \kappa_Y$) $\psi_T$ does not cancel and the equality is solved directly; when $\gamma_T\gamma_Y$ and $\kappa_T\kappa_Y$ have the same sign, as in the design, both population biases have that sign and the equality compares magnitudes, otherwise the absolute values are compared.

*Interaction with a covariate.* Let $g = UZ_1$ with $\rho = 0$. Then $g$ is centered and uncorrelated with $U$, the indicators, and every covariate ($E(U^2Z_1) = E(UZ_1Z_k) = 0$), so A.4 holds and SEM-L has the bias of Equation 5 with $\mathrm{Var}(g) = 1$. Because $Z_1$ is in every adjustment set, $E(g \mid X) = Z_1\mu$ and $h = Z_1 d$; since $d$ is independent of $X$, $E(dh) = E(Z_1)\tau^2 = 0$ and $V = E(Z_1^2)\tau^2 = \tau^2$, so Equation 4 applies. In the symmetric case the equality becomes $(\gamma^2 + \kappa^2)\tau^2 = \kappa^2$, that is, $\tau^{*2} = \pi$, and with parallel indicators $\omega^* = 1 - \pi$, lower than the quadratic crossover reliability at every $\pi$ (at $\omega = 1 - \pi$ the left side of $a\omega^2 + \omega - 1 = 0$ equals $-\pi^2 < 0$). A quadratic term captures none of $g$, because $E\{UZ_1(U^2 - 1)\} = 0$; an LMS model with the product term $UZ_1$ would be correctly specified.

### A.7 Coverage Under Persistent Bias

If an estimator is approximately $N(\beta + b, s_N^2)$ with a correct standard error, its coverage of $\beta$ is $\Phi(1.96 - b/s_N) - \Phi(-1.96 - b/s_N)$. For SEM-L under quadratic confounding ($b$ = 0.153) with the mean standard errors of Tables C1a and C1b (0.068–0.076 at $N$ = 250; 0.034–0.038 at $N$ = 1,000) this gives .39–.49 and .01–.02, against the observed .38–.52 and .00–.01; for SEM-Q under the saturating form ($b$ = 0.051; standard errors 0.033–0.038 at $N$ = 1,000) it gives .66–.74, against the observed .64–.72. Over all conditions and estimators with a population target (Table A1), the coverage of that target was .92–.97 for SEM-L, .94–.97 for SEM-Q under the linear and quadratic forms, .95–.97 for Reg-L and Reg-Q, and .92–.96 for proxy-input DML at $N$ = 1,000, except under the cubic form at $\omega$ = .93 (.86–.89); at $N$ = 250 it was .89–.95 for proxy-input DML and .89–.92 for DML-True, where the cost of learning displaces DML from its target. The intervals are therefore valid for the quantity each estimator converges to, and the coverage of $\beta$ fails through bias.

## A.8 Population Bias in Every Condition

**Table A1**

*Population Bias of Each Approach by Form of Confounding (a) at the Core Reliability Levels and (b) in the Crossover and Sensitivity Conditions*

(a) Core reliability levels (parallel indicators, ρ = 0)

| Form | SEM-L | DML, ω = .63 | DML, ω = .74 | DML, ω = .84 | DML, ω = .93 | Reg-Q | SEM-Q |
|---|---|---|---|---|---|---|---|
| Linear | 0 | 0.085 | 0.060 | 0.038 | 0.018 | 0 | 0 |
| Quadratic | 0.153 | 0.168 | 0.127 | 0.084 | 0.041 | 0 | 0 |
| Saturating | 0.153 | 0.179 | 0.140 | 0.098 | 0.051 | 0.053 | — |
| Cubic | 0.153 | 0.186 | 0.146 | 0.101 | 0.052 | 0.153 | — |

(b) Crossover and sensitivity conditions ($N$ = 1,000)

| Block | Form | ω | ρ | SEM-L | DML-Mean | DML-Items, DML-Opt | Reg-Q | SEM-Q |
|---|---|---|---|---|---|---|---|---|
| Crossover | Quadratic | .67 | 0 | 0.153 | 0.153 | 0.153 | 0 | 0 |
| Crossover | Saturating | .71 | 0 | 0.153 | 0.153 | 0.153 | 0.053 | — |
| Correlated covariates | Quadratic | .63 | .6 | 0.153 | 0.146 | 0.146 | 0 | 0 |
| Correlated covariates | Quadratic | .63 | .8 | 0.153 | 0.113 | 0.113 | 0 | 0 |
| Correlated covariates | Quadratic | .93 | .6 | 0.153 | 0.040 | 0.040 | 0 | 0 |
| Correlated covariates | Quadratic | .93 | .8 | 0.153 | 0.037 | 0.037 | 0 | 0 |
| Correlated covariates | Saturating | .63 | .6 | 0.153 | 0.158 | 0.158 | 0.053 | — |
| Correlated covariates | Saturating | .63 | .8 | 0.153 | 0.127 | 0.127 | 0.053 | — |
| Correlated covariates | Saturating | .93 | .6 | 0.153 | 0.049 | 0.049 | 0.053 | — |
| Correlated covariates | Saturating | .93 | .8 | 0.153 | 0.046 | 0.046 | 0.053 | — |
| Unequal loadings ($\omega_{\text{opt}}$ = .76) | Quadratic | .63 | 0 | 0.153 | 0.168 | 0.119 | 0 | 0 |
| Unequal loadings ($\omega_{\text{opt}}$ = .90) | Quadratic | .84 | 0 | 0.153 | 0.084 | 0.056 | 0 | 0 |
| Unequal loadings ($\omega_{\text{opt}}$ = .76) | Saturating | .63 | 0 | 0.153 | 0.179 | 0.132 | 0.053 | — |
| Unequal loadings ($\omega_{\text{opt}}$ = .90) | Saturating | .84 | 0 | 0.153 | 0.098 | 0.068 | 0.053 | — |
| Cubic form | Cubic | .63 | 0 | 0.153 | 0.186 | 0.186 | 0.153 | — |
| Cubic form | Cubic | .93 | 0 | 0.153 | 0.052 | 0.052 | 0.153 | — |

*Note.* From Equations 4, 5, and 7 with the design values of the main text ($\beta$ = 0.30; $\gamma_T = \gamma_Y$ = 0.50; $\kappa_T = \kappa_Y$ = 0.30, 1.603, and 0.173 for the quadratic, saturating, and cubic forms). In panel (a) DML = DML-Items and DML-Mean, which have the same population bias under parallel indicators; under unequal loadings DML-Items and DML-Opt share $\tau^2 = 1 - \omega_{\text{opt}}$ (A.2). The

SEM biases do not depend on $\rho$ or on the loading pattern (A.4). Reg-Q = quadratic regression with observed $U$; under the cubic form a quadratic term recovers none of the nonlinear variance, so Reg-Q keeps the full linear bias (A.5). — = SEM-Q is misspecified under this form; its population bias, approximated at $N$ = 20,000, is 0.045–0.046 (saturating) and 0.143–0.147 (cubic; Table C4).

## Supplement B. Estimation Protocol

### B.1 Conditions, Seeds, and Data

The design (Figure 1; Equations 1–3) has 46 conditions with 500 replications each and seed 20261009. Each dataset is a deterministic function of the seed, the replication index, and the condition: independent random streams generate the latent innovation, the ten covariate innovations, the two structural errors, and the three indicator errors, and a replication index shares these draws across conditions, so comparisons between conditions and estimators use common random numbers; the $N$ = 250 datasets are the first 250 rows of the $N$ = 1,000 datasets. Pilot, tuning, diagnostic, and large-sample checks use separate seed streams. With 46 free parameters in SEM-L and 48 in SEM-Q, $N$ = 250 gives about five observations per free parameter, the lower limit that Bentler and Chou (1987) suggested, and $N$ = 1,000 about 20.

Unequal loadings are $s \times (0.5,1,1.4)$ at $\omega$ = .63 and $s \times (0.8,1,1.15)$ at $\omega$ = .84, with $s$ chosen so that the composite reliability of the indicator mean matches the parallel counterpart: loadings .303/.605/.847 (error variances 0.908/0.634/0.282; $\omega_{\text{opt}}$ = .763) and .646/.808/.929 (error variances 0.582/0.347/0.136; $\omega_{\text{opt}}$ = .899); at $\omega$ = .84 stronger heterogeneity would need a loading above .93. The crossover block uses the common loading at which Equation 4 equals Equation 5 (A.6), .638 under quadratic and .669 under saturating confounding; the correlated-covariate block uses $\rho$ = .6 and .8 ($\rho$ = .3 adds only 0.10 to the information about $U$ and was not simulated); the cubic block uses common loadings .6 and .9.

### B.2 Mplus Models and the Scaled Difference Test

Both latent models fix the first loading to 1, estimate the other loadings and error variances, regress $U$ on $Z_1$–$Z_{10}$, and use MLR (Satorra & Bentler, 1994; Yuan & Bentler, 2000) with the Mplus defaults for convergence. Regressing the latent variable on observed covariates is equivalent, in the linear model, to estimating their covariances; the regression is used because Mplus treats observed covariates as exogenous. SEM-Q adds the latent product term with `TYPE = RANDOM`, `ALGORITHM = INTEGRATION`, and `INTEGRATION = STANDARD(30)`, and uses the converged SEM-L loadings of the same dataset as starting values. The SEM-Q syntax, apart from the TITLE and DATA commands, is

```
VARIABLE:
    NAMES ARE u1 u2 u3 Z1 Z2 Z3 Z4 Z5 Z6 Z7 Z8 Z9 Z10 T Y;
    USEVARIABLES ARE u1 u2 u3 Z1 Z2 Z3 Z4 Z5 Z6 Z7 Z8 Z9 Z10 T
Y;
ANALYSIS:
    ESTIMATOR = MLR;
    TYPE = RANDOM;
    ALGORITHM = INTEGRATION;
    INTEGRATION = STANDARD (30);
MODEL:
```

```
    U BY u1@1 u2*<SEM-L estimate> u3*<SEM-L estimate>;
    U ON Z1 Z2 Z3 Z4 Z5 Z6 Z7 Z8 Z9 Z10;
    UQ | U XWITH U;
    T ON U UQ Z1 Z2 Z3 Z4 Z5 Z6 Z7 Z8 Z9 Z10;
    Y ON T (bt);
    Y ON U UQ Z1 Z2 Z3 Z4 Z5 Z6 Z7 Z8 Z9 Z10;
MODEL CONSTRAINT:
    NEW(btprec);
    btprec = 1000*bt;
OUTPUT: TECH1 CINTERVAL;
```

SEM-L omits the `TYPE = RANDOM` lines, `UQ`, and the starting values. LMS (Klein & Moosbrugger, 2000) assumes a normally distributed latent predictor; because $U$ is regressed on $\boldsymbol{Z}$, the assumption applies to the residual of $U$, which is normal in every condition. The derived parameter `btprec`, the treatment coefficient multiplied by 1,000, is read with three decimals and so yields six decimals of the coefficient and its standard error. Time limits are 300 s (SEM-L) and 1,200 s (SEM-Q) per fit.

SEM-L and SEM-Q are competing models specified in advance, the alternative-models strategy of Jöreskog (1993), so the choice between them is not a specification search (MacCallum, 1986). SEM-Select uses the scaled difference test (Satorra & Bentler, 2001) in the loglikelihood form that Mplus uses for MLR: with $T_k = -2$ loglikelihood, $c_k$ the scaling correction factor, and $p_k$ the number of free parameters of model $k$ (0 = SEM-L, 1 = SEM-Q), $\bar{T} = (T_0 - T_1)/c_d$ with $c_d = (p_0 c_0 - p_1 c_1)/(p_0 - p_1)$ is referred to $\chi^2(2)$, and SEM-Q is chosen when $p < .05$. When SEM-Q is unusable or $c_d \leq 0$ (Satorra & Bentler, 2010), SEM-Select reports SEM-L and records the reason; SEM-Q was usable in every replication.

With $p_0 = 46$ and $p_1 = 48$, $c_d = 24c_1 - 23c_0$: it is nonpositive whenever $c_0$ exceeds $c_1$ by more than about 4%, and otherwise it amplifies the difference between the two corrections about 23-fold, so the scaled statistic is unstable whenever the two corrections are close. Under quadratic confounding the omitted term inflates $c_0$ (mean 1.02–1.03) while $c_1$ of the correctly specified SEM-Q stays near 1 (0.99–1.00); this is why the statistic was undefined in 5–15% of the quadratic replications and in at most 0.2% under the other forms (Table C3). Under the cubic form both models are misspecified and $c_1$ (1.21–1.24) exceeds $c_0$ (1.08–1.09), so $c_d$ averaged 4.7 ($\omega$ = .63) and 4.2 ($\omega$ = .93) and the scaled statistic was a quarter to a fifth of the unscaled one, which explains part of the low rate at which the test detected this form. In an exploratory check that was not preregistered, the unscaled statistic $T_0 - T_1$ rejected in all 726 replications with an undefined scaled statistic, so a rule that fell back on it would have selected SEM-Q in 99–100% of the quadratic replications; the check changes no reported estimate. The strictly positive version of the test (Satorra & Bentler, 2010; in Mplus, Asparouhov & Muthén, 2010) and a Wald test of the two quadratic coefficients (MODEL TEST in Mplus) avoid the undefined case; neither was part of the preregistered protocol. The interval that SEM-Select reports ignores that the model was chosen on the same data (MacCallum et al., 1992; Leeb & Pötscher, 2005).

The nonnormality of $T$ and $Y$ is mild except under the cubic form: with the design values their population skewness and excess kurtosis are 0.37/0.51 and 0.68/1.02 (quadratic), 0.22/0.30 and 0.10/0.15 (saturating), and 0/0 and 3.9/5.9 (cubic), all below the values of 2 and 7 at which Curran et al. (1996) found maximum-likelihood test statistics to be seriously affected. Under the excess kurtosis of the cubic form, the mean standard errors were up to 11% too small in those two conditions, for the DML estimators as well as for the latent models (Table C6d): a finite-sample shortfall of sandwich-type standard errors under heavy-tailed data, which does not affect the comparison of biases. Fits of SEM-Q to three datasets of $N$ = 20,000 per misspecified condition with 30 and 60 integration points confirm that the integration setting does not affect the estimates (Table C4).

**B.3 DML**

All DML variants use the partially linear partialling-out score of DoubleML 0.11.4 with five-fold cross-fitting, one sample split, and the confidence limits it returns; the nuisance learner is the histogram-based gradient-boosting regressor of scikit-learn 1.8.0 without early stopping. Hyperparameters are selected once per condition and estimator (*pretuning*): 24 candidates are drawn with a fixed seed from the grid in Table B1 (36 combinations at $N$ = 250, 108 at $N$ = 1,000) and scored by three-fold cross-validated mean squared prediction error of $T$ and, separately, of $Y$ (a criterion that Bach et al., 2024, found to work well) on five independent datasets from the tuning seed stream; the candidate with the best mean score is then held fixed for the 500 replications. Pretuning uses datasets from the same condition, which an applied analyst would not have; it replaces a nested search inside every training fold of every replication. Of the 204 selections, 192 used the smallest learning rate (0.01) and 166 the smallest maximum depth of the grid, and at $N$ = 1,000, 107 of 150 also used the largest number of iterations (500); the reported cost of learning is that of this grid.

**Table B1**
*Hyperparameter Grid*

| Setting | $N$ = 250 | $N$ = 1,000 |
|---|---|---|
| Learning rate | 0.01, 0.05, 0.10 | 0.01, 0.05, 0.10, 0.20 |
| Maximum depth | 2, 3 | 3, 4, 5 |
| Minimum samples per leaf | 5, 10, 20 | 10, 20, 50 |
| Boosting iterations | 100, 300 | 100, 300, 500 |

*Exploratory check with a wider grid (not preregistered; changes no reported estimate).* Because most selected settings lay at the boundary of Table B1, DML-Mean was refitted in three conditions (quadratic at $\omega^*$, $N$ = 1,000; quadratic at $\omega$ = .93, $N$ = 250; cubic at $\omega$ = .93) with hyperparameters selected by the same protocol from a wider grid (48 random candidates; at $N$ = 1,000 learning rates 0.001–0.2, 100–2,000 iterations, depths 2–5, and 10–100 observations per leaf). Cross-validation again chose settings at an edge of the grid, now the shallowest trees (depth 2 in five of the six selections), and the cost of learning did not fall: 0.012 against 0.005 in the main run, 0.046 against 0.046, and 0.052 against 0.045 (Monte Carlo standard errors of the differences 0.002–0.004). The settings that minimize prediction error are therefore not those that minimize the bias of the orthogonal score (Bach et al., 2024).

DML-FS estimates a one-factor model by maximum likelihood on the three indicators of each training fold and computes regression factor scores for the validation fold from that model, so that no validation data enter the score; a fold whose measurement model does not converge or reaches a zero error variance counts as a failed fit. This happened only in the two unequal-loadings conditions at $\omega$ = .63 (loadings .30/.61/.85), in the same 36 of 500 replications under both forms (the indicators are identical across forms). In every failed fold the sample covariances implied a negative error variance for the strongest indicator (a Heywood case): with three indicators the one-factor model is just-identified (Bollen, 1989), so the unrestricted solution for the squared loading of indicator 3 is $s_{13}s_{23}/s_{12}$, which exceeds $s_{33}$ when $s_{12}$ is small by chance, and the iterations then approach the boundary $\theta_3 = 0$ (Anderson & Gerbing, 1984). The failures reflect sampling error in a correctly specified, just-identified measurement model (a diagnosis that was not preregistered). DML-Opt uses the known weights $\lambda_j/\theta_j$ and is an oracle benchmark.

### B.4 Observed-Confounder Regressions and the Check of Equation 4

Reg-L and Reg-Q are ordinary least squares of $Y$ on $T$, $U$ (and $U^2$), and $Z_1$–$Z_{10}$ with HC3 standard errors (MacKinnon & White, 1985); DML-True applies the DML protocol with $U$ in place of the indicators. Equation 4 was checked by evaluating the partialling-out score with the true conditional means $E(T \mid X)$ and $E(Y \mid X)$ on five datasets of $N$ = 20,000 for each of the 31 conditions with $N$ = 1,000: every deviation from Equation 4 was at most 0.006 in absolute value and within about two Monte Carlo standard errors. The cost of learning is measured in the main run as the DML bias minus Equation 4; the Monte Carlo standard error of this difference is about 0.0015 at $N$ = 1,000 and 0.003 at $N$ = 250.

### B.5 Failure Accounting

Every attempted fit receives exactly one status: usable (terminated normally with a proper solution and a finite estimate and standard error), nonconverged, improper (negative variance, nonpositive-definite matrix, or standard errors that could not be computed), timeout, numerical error, or execution error. Performance is summarized over usable fits; a usable Mplus fit whose standard errors carry the warning that they may not be trustworthy keeps its estimate but is excluded from the standard-error and coverage summaries. Numerical failures stay in the denominator and are never replaced, repaired, or retried; execution errors are retried, with all attempts retained. Paired differences between estimators are computed over the replications usable for both. In the main run (97,000 attempted fits), 72 fits did not converge, all of them DML-FS fits in the two unequal-loadings conditions at $\omega$ = .63 (B.3); no fit was improper, reached a time limit, or carried the standard-error warning, and no execution error remained.

## Supplement C. Results

Tables C1–C3 and C6 and Figure C1 report the main run; Table C4 reports the integration check at $N$ = 20,000 (B.2). Every SEM-L, SEM-Q, DML-Items, DML-Mean, DML-Opt, Reg-L, Reg-Q, and DML-True fit converged to a proper solution; the convergence rate is therefore shown only in Table C6c, where DML-FS failed (B.3).

**Table C1a**

*Core Conditions, N = 1,000: Bias, Relative Bias, Empirical SD, Mean Standard Error, RMSE, and 95% Interval Coverage (500 Replications)*

| Form | ω | Estimator | Bias | RB (%) | SD | Mean SE | RMSE | Coverage |
|---|---|---|---|---|---|---|---|---|
| Linear | .63 | SEM-L | 0.000 | 0.0 | 0.036 | 0.037 | 0.036 | .966 |

| Form | ω | Estimator | Bias | RB (%) | SD | Mean SE | RMSE | Coverage |
|---|---|---|---|---|---|---|---|---|
| | | SEM-Q | 0.000 | 0.0 | 0.036 | 0.037 | 0.036 | .972 |
| | | DML-Items | 0.092 | 30.6 | 0.033 | 0.032 | 0.098 | .200 |
| | | DML-Mean | 0.086 | 28.6 | 0.033 | 0.032 | 0.092 | .240 |
| | | SEM-Select | 0.000 | 0.0 | 0.036 | 0.037 | 0.036 | .966 |
| Linear | .74 | SEM-L | 0.001 | 0.4 | 0.034 | 0.035 | 0.034 | .966 |
| | | SEM-Q | 0.001 | 0.3 | 0.034 | 0.035 | 0.034 | .966 |
| | | DML-Items | 0.067 | 22.2 | 0.033 | 0.032 | 0.074 | .446 |
| | | DML-Mean | 0.063 | 20.8 | 0.033 | 0.032 | 0.071 | .500 |
| | | SEM-Select | 0.001 | 0.4 | 0.034 | 0.035 | 0.034 | .964 |
| Linear | .84 | SEM-L | 0.002 | 0.6 | 0.033 | 0.033 | 0.033 | .964 |
| | | SEM-Q | 0.002 | 0.5 | 0.033 | 0.033 | 0.033 | .960 |
| | | DML-Items | 0.041 | 13.6 | 0.033 | 0.032 | 0.052 | .740 |
| | | DML-Mean | 0.041 | 13.7 | 0.033 | 0.032 | 0.052 | .746 |
| | | SEM-Select | 0.002 | 0.6 | 0.033 | 0.033 | 0.033 | .964 |
| Linear | .93 | SEM-L | 0.002 | 0.7 | 0.032 | 0.032 | 0.032 | .960 |
| | | SEM-Q | 0.002 | 0.7 | 0.032 | 0.032 | 0.032 | .958 |
| | | DML-Items | 0.025 | 8.2 | 0.032 | 0.032 | 0.041 | .876 |
| | | DML-Mean | 0.026 | 8.6 | 0.033 | 0.032 | 0.042 | .864 |
| | | SEM-Select | 0.002 | 0.7 | 0.032 | 0.032 | 0.032 | .960 |
| Quadratic | .63 | SEM-L | 0.152 | 50.6 | 0.036 | 0.038 | 0.156 | .014 |
| | | SEM-Q | 0.001 | 0.2 | 0.039 | 0.039 | 0.039 | .946 |
| | | DML-Items | 0.176 | 58.5 | 0.034 | 0.034 | 0.179 | .000 |
| | | DML-Mean | 0.171 | 57.1 | 0.035 | 0.034 | 0.175 | .002 |
| | | SEM-Select | 0.010 | 3.2 | 0.053 | 0.039 | 0.054 | .900 |
| Quadratic | .67 | SEM-L | 0.152 | 50.7 | 0.035 | 0.037 | 0.156 | .010 |
| | | SEM-Q | 0.000 | 0.2 | 0.038 | 0.038 | 0.038 | .946 |
| | | DML-Items | 0.161 | 53.7 | 0.034 | 0.034 | 0.165 | .000 |
| | | DML-Mean | 0.157 | 52.5 | 0.035 | 0.034 | 0.161 | .002 |
| | | SEM-Select | 0.010 | 3.4 | 0.053 | 0.038 | 0.053 | .898 |
| Quadratic | .74 | SEM-L | 0.152 | 50.8 | 0.034 | 0.036 | 0.156 | .008 |
| | | SEM-Q | 0.001 | 0.2 | 0.037 | 0.037 | 0.037 | .954 |
| | | DML-Items | 0.136 | 45.5 | 0.034 | 0.033 | 0.141 | .014 |
| | | DML-Mean | 0.134 | 44.6 | 0.034 | 0.033 | 0.138 | .016 |
| | | SEM-Select | 0.011 | 3.8 | 0.052 | 0.037 | 0.054 | .900 |
| Quadratic | .84 | SEM-L | 0.153 | 51.0 | 0.033 | 0.035 | 0.157 | .006 |

| Form | ω | Estimator | Bias | RB (%) | SD | Mean SE | RMSE | Coverage |
|---|---|---|---|---|---|---|---|---|
| | | SEM-Q | 0.001 | 0.4 | 0.034 | 0.035 | 0.034 | .962 |
| | | DML-Items | 0.095 | 31.8 | 0.033 | 0.033 | 0.101 | .184 |
| | | DML-Mean | 0.094 | 31.4 | 0.033 | 0.033 | 0.100 | .200 |
| | | SEM-Select | 0.015 | 4.9 | 0.054 | 0.035 | 0.056 | .892 |
| Quadratic | .93 | SEM-L | 0.153 | 51.1 | 0.033 | 0.034 | 0.157 | .000 |
| | | SEM-Q | 0.002 | 0.5 | 0.033 | 0.033 | 0.033 | .962 |
| | | DML-Items | 0.055 | 18.2 | 0.032 | 0.033 | 0.063 | .632 |
| | | DML-Mean | 0.054 | 18.1 | 0.033 | 0.033 | 0.063 | .618 |
| | | SEM-Select | 0.022 | 7.4 | 0.062 | 0.034 | 0.065 | .848 |
| Saturating | .63 | SEM-L | 0.153 | 50.8 | 0.034 | 0.035 | 0.156 | .006 |
| | | SEM-Q | 0.051 | 17.0 | 0.038 | 0.038 | 0.064 | .724 |
| | | DML-Items | 0.182 | 60.7 | 0.033 | 0.032 | 0.185 | .000 |
| | | DML-Mean | 0.181 | 60.4 | 0.034 | 0.032 | 0.184 | .000 |
| | | SEM-Select | 0.051 | 17.0 | 0.038 | 0.038 | 0.064 | .724 |
| Saturating | .71 | SEM-L | 0.153 | 51.0 | 0.033 | 0.034 | 0.157 | .004 |
| | | SEM-Q | 0.051 | 16.9 | 0.037 | 0.037 | 0.063 | .706 |
| | | DML-Items | 0.157 | 52.4 | 0.033 | 0.032 | 0.161 | .000 |
| | | DML-Mean | 0.155 | 51.8 | 0.034 | 0.032 | 0.159 | .000 |
| | | SEM-Select | 0.051 | 16.9 | 0.037 | 0.037 | 0.063 | .706 |
| Saturating | .74 | SEM-L | 0.153 | 51.1 | 0.032 | 0.033 | 0.157 | .004 |
| | | SEM-Q | 0.051 | 16.8 | 0.036 | 0.036 | 0.062 | .706 |
| | | DML-Items | 0.150 | 50.2 | 0.033 | 0.032 | 0.154 | .002 |
| | | DML-Mean | 0.144 | 48.0 | 0.033 | 0.032 | 0.148 | .004 |
| | | SEM-Select | 0.051 | 16.8 | 0.036 | 0.036 | 0.062 | .706 |
| Saturating | .84 | SEM-L | 0.154 | 51.2 | 0.032 | 0.032 | 0.157 | .002 |
| | | SEM-Q | 0.050 | 16.8 | 0.034 | 0.034 | 0.061 | .686 |
| | | DML-Items | 0.109 | 36.5 | 0.034 | 0.032 | 0.114 | .080 |
| | | DML-Mean | 0.103 | 34.5 | 0.033 | 0.032 | 0.108 | .120 |
| | | SEM-Select | 0.050 | 16.8 | 0.034 | 0.034 | 0.061 | .686 |
| Saturating | .93 | SEM-L | 0.154 | 51.3 | 0.031 | 0.031 | 0.157 | .002 |
| | | SEM-Q | 0.051 | 17.0 | 0.033 | 0.033 | 0.060 | .642 |
| | | DML-Items | 0.064 | 21.3 | 0.033 | 0.032 | 0.072 | .510 |
| | | DML-Mean | 0.059 | 19.8 | 0.033 | 0.032 | 0.068 | .546 |
| | | SEM-Select | 0.051 | 17.0 | 0.033 | 0.033 | 0.060 | .642 |

*Note.* RB = relative bias (bias / 0.30). Rows at ω = .67 (quadratic) and .71 (saturating) are the crossover conditions. Coverage = proportion of replications whose 95% interval contains β = 0.30; the Monte Carlo standard error of a coverage of .95 is .010. SEM-Select reports SEM-Q when the scaled difference test rejects at α = .05 and SEM-L otherwise, including when the test is undefined (rates in Table C3).

**Table C1b**

*Core Conditions, N = 250: Bias, Relative Bias, Empirical SD, Mean Standard Error, RMSE, and 95% Interval Coverage (500 Replications)*

| Form | ω | Estimator | Bias | RB (%) | SD | Mean SE | RMSE | Coverage |
|---|---|---|---|---|---|---|---|---|
| Linear | .63 | SEM-L | −0.006 | −2.0 | 0.074 | 0.077 | 0.074 | .956 |
| | | SEM-Q | −0.007 | −2.5 | 0.076 | 0.079 | 0.076 | .964 |
| | | DML-Items | 0.120 | 40.1 | 0.061 | 0.064 | 0.135 | .528 |
| | | DML-Mean | 0.115 | 38.3 | 0.061 | 0.064 | 0.130 | .544 |
| | | SEM-Select | −0.006 | −2.0 | 0.074 | 0.077 | 0.074 | .958 |
| Linear | .74 | SEM-L | −0.002 | −0.6 | 0.068 | 0.070 | 0.068 | .948 |
| | | SEM-Q | −0.002 | −0.8 | 0.069 | 0.071 | 0.069 | .950 |
| | | DML-Items | 0.077 | 25.5 | 0.062 | 0.064 | 0.099 | .774 |
| | | DML-Mean | 0.096 | 32.0 | 0.062 | 0.064 | 0.114 | .686 |
| | | SEM-Select | −0.002 | −0.6 | 0.068 | 0.070 | 0.068 | .950 |
| Linear | .84 | SEM-L | 0.001 | 0.2 | 0.065 | 0.067 | 0.065 | .948 |
| | | SEM-Q | 0.000 | 0.1 | 0.065 | 0.067 | 0.065 | .954 |
| | | DML-Items | 0.075 | 24.9 | 0.061 | 0.064 | 0.097 | .796 |
| | | DML-Mean | 0.075 | 25.0 | 0.062 | 0.064 | 0.098 | .792 |
| | | SEM-Select | 0.001 | 0.3 | 0.065 | 0.067 | 0.065 | .950 |
| Linear | .93 | SEM-L | 0.002 | 0.7 | 0.063 | 0.065 | 0.063 | .950 |
| | | SEM-Q | 0.002 | 0.7 | 0.063 | 0.065 | 0.063 | .950 |
| | | DML-Items | 0.054 | 18.1 | 0.062 | 0.064 | 0.082 | .878 |
| | | DML-Mean | 0.040 | 13.5 | 0.063 | 0.064 | 0.075 | .912 |
| | | SEM-Select | 0.002 | 0.7 | 0.063 | 0.065 | 0.063 | .948 |
| Quadratic | .63 | SEM-L | 0.147 | 48.9 | 0.071 | 0.076 | 0.163 | .520 |
| | | SEM-Q | −0.007 | −2.3 | 0.083 | 0.082 | 0.084 | .948 |
| | | DML-Items | 0.193 | 64.3 | 0.069 | 0.069 | 0.205 | .190 |
| | | DML-Mean | 0.203 | 67.6 | 0.064 | 0.069 | 0.213 | .144 |
| | | SEM-Select | 0.023 | 7.7 | 0.098 | 0.082 | 0.101 | .884 |
| Quadratic | .74 | SEM-L | 0.150 | 49.9 | 0.068 | 0.072 | 0.164 | .452 |
| | | SEM-Q | −0.004 | −1.3 | 0.076 | 0.075 | 0.076 | .950 |

| Form | ω | Estimator | Bias | RB (%) | SD | Mean SE | RMSE | Coverage |
|---|---|---|---|---|---|---|---|---|
| | | DML-Items | 0.157 | 52.2 | 0.068 | 0.068 | 0.171 | .344 |
| | | DML-Mean | 0.155 | 51.8 | 0.070 | 0.068 | 0.170 | .356 |
| | | SEM-Select | 0.023 | 7.6 | 0.094 | 0.075 | 0.096 | .880 |
| Quadratic | .84 | SEM-L | 0.152 | 50.5 | 0.066 | 0.069 | 0.165 | .402 |
| | | SEM-Q | −0.001 | −0.3 | 0.070 | 0.070 | 0.070 | .946 |
| | | DML-Items | 0.124 | 41.4 | 0.068 | 0.068 | 0.142 | .550 |
| | | DML-Mean | 0.122 | 40.6 | 0.069 | 0.068 | 0.140 | .570 |
| | | SEM-Select | 0.026 | 8.5 | 0.091 | 0.071 | 0.094 | .864 |
| Quadratic | .93 | SEM-L | 0.153 | 51.0 | 0.065 | 0.068 | 0.166 | .380 |
| | | SEM-Q | 0.001 | 0.4 | 0.065 | 0.066 | 0.065 | .944 |
| | | DML-Items | 0.075 | 25.0 | 0.070 | 0.067 | 0.102 | .810 |
| | | DML-Mean | 0.087 | 29.1 | 0.069 | 0.068 | 0.111 | .760 |
| | | SEM-Select | 0.030 | 10.2 | 0.091 | 0.067 | 0.096 | .846 |
| Saturating | .63 | SEM-L | 0.148 | 49.2 | 0.067 | 0.072 | 0.162 | .460 |
| | | SEM-Q | 0.043 | 14.2 | 0.081 | 0.081 | 0.091 | .898 |
| | | DML-Items | 0.215 | 71.8 | 0.060 | 0.064 | 0.224 | .054 |
| | | DML-Mean | 0.204 | 67.9 | 0.061 | 0.064 | 0.212 | .092 |
| | | SEM-Select | 0.044 | 14.6 | 0.081 | 0.081 | 0.092 | .896 |
| Saturating | .74 | SEM-L | 0.151 | 50.2 | 0.063 | 0.067 | 0.163 | .390 |
| | | SEM-Q | 0.045 | 15.1 | 0.074 | 0.074 | 0.086 | .904 |
| | | DML-Items | 0.165 | 55.1 | 0.062 | 0.064 | 0.176 | .264 |
| | | DML-Mean | 0.171 | 57.1 | 0.061 | 0.064 | 0.182 | .224 |
| | | SEM-Select | 0.045 | 15.2 | 0.074 | 0.074 | 0.087 | .904 |
| Saturating | .84 | SEM-L | 0.153 | 50.9 | 0.061 | 0.065 | 0.164 | .324 |
| | | SEM-Q | 0.047 | 15.7 | 0.069 | 0.069 | 0.083 | .890 |
| | | DML-Items | 0.139 | 46.4 | 0.061 | 0.064 | 0.152 | .414 |
| | | DML-Mean | 0.117 | 39.1 | 0.067 | 0.065 | 0.135 | .554 |
| | | SEM-Select | 0.047 | 15.7 | 0.069 | 0.069 | 0.083 | .888 |
| Saturating | .93 | SEM-L | 0.154 | 51.3 | 0.060 | 0.063 | 0.165 | .300 |
| | | SEM-Q | 0.049 | 16.3 | 0.065 | 0.066 | 0.082 | .878 |
| | | DML-Items | 0.080 | 26.6 | 0.066 | 0.065 | 0.103 | .758 |
| | | DML-Mean | 0.089 | 29.7 | 0.063 | 0.064 | 0.109 | .716 |
| | | SEM-Select | 0.049 | 16.3 | 0.065 | 0.066 | 0.082 | .876 |

*Note.* See Table C1a for column definitions.

**Table C2**

*SEM-L Fit Statistics in the Core Conditions at N = 1,000 (Means Over 500 Replications)*

| Form | *N* | ω | $\chi^2$ (*df* = 24) | TLI | CFI | RMSEA | SRMR | Conventional (%) | Hu–Bentler (%) | $\chi^2$ rejected (%) |
|---|---|---|---|---|---|---|---|---|---|---|
| Linear | 1,000 | .63 | 24.36 | .993 | .997 | .007 | .014 | 100.0 | 99.2 | 5.6 |
| Linear | 1,000 | .74 | 24.30 | .995 | .998 | .007 | .013 | 100.0 | 99.8 | 5.6 |
| Linear | 1,000 | .84 | 24.26 | .996 | .999 | .007 | .011 | 100.0 | 100.0 | 5.2 |
| Linear | 1,000 | .93 | 24.23 | .998 | .999 | .007 | .008 | 100.0 | 100.0 | 5.2 |
| Quadratic | 1,000 | .63 | 24.33 | .993 | .997 | .007 | .014 | 100.0 | 99.2 | 5.8 |
| Quadratic | 1,000 | .74 | 24.28 | .995 | .998 | .007 | .013 | 100.0 | 99.8 | 6.0 |
| Quadratic | 1,000 | .84 | 24.25 | .996 | .999 | .007 | .011 | 100.0 | 100.0 | 5.8 |
| Quadratic | 1,000 | .93 | 24.23 | .998 | .999 | .007 | .008 | 100.0 | 100.0 | 5.6 |
| Saturating | 1,000 | .63 | 24.33 | .993 | .997 | .007 | .014 | 100.0 | 99.4 | 5.4 |
| Saturating | 1,000 | .74 | 24.28 | .995 | .998 | .007 | .013 | 100.0 | 99.8 | 5.6 |
| Saturating | 1,000 | .84 | 24.24 | .996 | .999 | .007 | .011 | 100.0 | 100.0 | 5.6 |
| Saturating | 1,000 | .93 | 24.21 | .998 | .999 | .007 | .008 | 100.0 | 100.0 | 5.4 |

*Note.* TLI = Tucker–Lewis index; CFI = comparative fit index; RMSEA = root mean square error of approximation; SRMR = standardized root mean square residual. LMS provides no fit indices, so fit statistics are reported for SEM-L only. Conventional = TLI and CFI ≥ .90 and RMSEA ≤ .08 (Bentler, 1990; Bentler & Bonett, 1980; Browne & Cudeck, 1993; Tucker & Lewis, 1973); Hu–Bentler = TLI and CFI ≥ .95, RMSEA ≤ .06, and SRMR ≤ .08 (Hu & Bentler, 1999); $\chi^2$ rejected = percentage of MLR chi-square tests with $p < .05$. The means do not depend on the form of confounding, as A.4 implies: the linear model reproduces the population covariance matrix under every form. At $N = 250$ the means in the 12 core conditions were $\chi^2$ 25.40–25.77, TLI .961–.989, CFI .984–.995, RMSEA .016–.017, and SRMR .016–.029; the conventional cutoffs were met in 86.2–100%, the Hu–Bentler cutoffs in 70.2–96.2%, and $\chi^2$ was rejected in 6.6–10.0%. In the two crossover conditions and the 14 sensitivity conditions with a latent confounder (all at $N = 1{,}000$) the means were $\chi^2$ 24.08–24.34, TLI .993–.998, CFI .997–.999, RMSEA .007, and SRMR .008–.015; the conventional cutoffs were met in 100%, the Hu–Bentler cutoffs in 99.0–100%, and $\chi^2$ was rejected in 4.8–6.0%.

**Table C3**

*SEM-Select: Percentage of Replications in Which SEM-Q Was Selected and in Which the Scaled Statistic Was Undefined, Core, Crossover, and Cubic Conditions (500 Replications per Condition)*

| Block | Form | *N* | ω | Selected SEM-Q (%) | Undefined (%) |
|---|---|---|---|---|---|
| Core | Linear | 250 | .63 | 6.6 | 0.0 |
| Core | Linear | 250 | .74 | 7.0 | 0.0 |
| Core | Linear | 250 | .84 | 6.6 | 0.0 |

| Block | Form | *N* | ω | Selected SEM-Q (%) | Undefined (%) |
|---|---|---|---|---|---|
| Core | Linear | 250 | .93 | 7.8 | 0.0 |
| Core | Linear | 1,000 | .63 | 5.0 | 0.0 |
| Core | Linear | 1,000 | .74 | 5.0 | 0.0 |
| Core | Linear | 1,000 | .84 | 4.6 | 0.0 |
| Core | Linear | 1,000 | .93 | 6.0 | 0.0 |
| Core | Quadratic | 250 | .63 | 86.2 | 13.0 |
| Core | Quadratic | 250 | .74 | 87.6 | 12.4 |
| Core | Quadratic | 250 | .84 | 87.0 | 13.0 |
| Core | Quadratic | 250 | .93 | 85.0 | 15.0 |
| Core | Quadratic | 1,000 | .63 | 95.2 | 4.8 |
| Core | Quadratic | 1,000 | .74 | 94.2 | 5.8 |
| Core | Quadratic | 1,000 | .84 | 92.4 | 7.6 |
| Core | Quadratic | 1,000 | .93 | 88.0 | 12.0 |
| Core | Saturating | 250 | .63 | 98.2 | 0.2 |
| Core | Saturating | 250 | .74 | 99.6 | 0.0 |
| Core | Saturating | 250 | .84 | 99.8 | 0.0 |
| Core | Saturating | 250 | .93 | 99.8 | 0.0 |
| Core | Saturating | 1,000 | .63 | 100.0 | 0.0 |
| Core | Saturating | 1,000 | .74 | 100.0 | 0.0 |
| Core | Saturating | 1,000 | .84 | 100.0 | 0.0 |
| Core | Saturating | 1,000 | .93 | 100.0 | 0.0 |
| Crossover | Quadratic | 1,000 | .67 | 94.8 | 5.2 |
| Crossover | Saturating | 1,000 | .71 | 100.0 | 0.0 |
| Cubic form | Cubic | 1,000 | .63 | 29.2 | 4.0 |
| Cubic form | Cubic | 1,000 | .93 | 13.8 | 0.2 |

*Note.* Selected = the scaled difference test (Satorra & Bentler, 2001; $df = 2$, $\alpha = .05$) rejected and SEM-Q was reported. Undefined = the difference scaling factor $c_d$ was not positive, so that the scaled statistic is undefined (Satorra & Bentler, 2010) and SEM-L was reported under the preregistered rule (B.2). In the 12 sensitivity conditions (correlated covariates and unequal loadings; $N = 1{,}000$) SEM-Q was selected in 88.4–95.4% of the quadratic replications (undefined in 4.6–11.6%) and in 100% of the saturating replications (undefined in 0%). No fallback was due to an unusable SEM-Q fit. A negative scaled statistic, which also selects SEM-L, occurred in 10 of the 20,000 replications (at most 4 per condition; linear and cubic forms). Under quadratic confounding the test rejected in more than 99% of the replications in which it was defined: selected and undefined sum to 100% except at $N = 250$, $\omega = .63$, where a defined test did not reject in 4 replications.

**Table C4**

*SEM-Q at N = 20,000 Under the Saturating and Cubic Forms (Three Datasets per Condition)*

| Form | ω | Mean estimate | Bias | Mean SE | Max. difference, 30 vs. 60 points |
|---|---|---|---|---|---|
| Saturating | .63 | 0.3464 | 0.046 | 0.0086 | 0.00001 |
| Saturating | .74 | 0.3459 | 0.046 | 0.0081 | 0.00003 |
| Saturating | .84 | 0.3453 | 0.045 | 0.0077 | 0.00000 |
| Saturating | .93 | 0.3460 | 0.046 | 0.0074 | 0.00000 |
| Cubic | .63 | 0.4433 | 0.143 | 0.0112 | 0.00000 |
| Cubic | .93 | 0.4467 | 0.147 | 0.0102 | 0.00000 |

*Note.* Separate seed stream; every fit terminated normally with a proper solution. Max. difference = largest absolute difference between the estimates with 30 and with 60 integration points. The population bias of SEM-Q under the saturating form is about 0.046 at every reliability, close to the 0.053 of Reg-Q with observed U (A.5); under the cubic form it is 0.143–0.147, within 0.01 of the SEM-L bias 0.153. At $N$ = 1,000 the SEM-Q bias was 0.050–0.051 under the saturating form and 0.128 and 0.145 under the cubic form at ω = .63 and .93 (Tables C1a and C6d). The saturating difference (about 0.005) is within one standard error of the three-dataset mean; the cubic difference at ω = .63 (0.015, about 2.3 standard errors) suggests that a misspecified LMS model can also carry a finite-sample bias (cf. Brandt et al., 2020, on LMS under structural misspecification).

**Table C6**

*Sensitivity Conditions*

(a) Observed confounder (no measurement error)

| Form | *N* | Estimator | Bias | RB (%) | SD | Mean SE | RMSE | Coverage |
|---|---|---|---|---|---|---|---|---|
| Linear | 250 | Reg-L | 0.003 | 1.0 | 0.063 | 0.067 | 0.063 | .964 |
| | | Reg-Q | 0.003 | 0.9 | 0.063 | 0.068 | 0.063 | .966 |
| | | DML-True | 0.041 | 13.8 | 0.064 | 0.064 | 0.076 | .894 |
| Linear | 1,000 | Reg-L | 0.002 | 0.8 | 0.032 | 0.032 | 0.032 | .962 |
| | | Reg-Q | 0.002 | 0.7 | 0.032 | 0.032 | 0.032 | .960 |
| | | DML-True | 0.009 | 2.9 | 0.033 | 0.032 | 0.034 | .942 |
| Quadratic | 250 | Reg-L | 0.154 | 51.3 | 0.066 | 0.072 | 0.167 | .412 |
| | | Reg-Q | 0.003 | 0.9 | 0.063 | 0.068 | 0.063 | .966 |
| | | DML-True | 0.043 | 14.4 | 0.072 | 0.068 | 0.084 | .894 |
| Quadratic | 1,000 | Reg-L | 0.153 | 51.2 | 0.033 | 0.035 | 0.157 | .000 |
| | | Reg-Q | 0.002 | 0.7 | 0.032 | 0.032 | 0.032 | .960 |
| | | DML-True | 0.017 | 5.7 | 0.033 | 0.033 | 0.037 | .914 |
| Saturating | 250 | Reg-L | 0.155 | 51.7 | 0.060 | 0.066 | 0.166 | .336 |

| Form | $N$ | Estimator | Bias | RB (%) | SD | Mean SE | RMSE | Coverage |
|---|---|---|---|---|---|---|---|---|
| | | Reg-Q | 0.053 | 17.6 | 0.063 | 0.068 | 0.082 | .894 |
| | | DML-True | 0.031 | 10.2 | 0.067 | 0.065 | 0.074 | .922 |
| Saturating | 1,000 | Reg-L | 0.154 | 51.4 | 0.031 | 0.031 | 0.157 | .000 |
| | | Reg-Q | 0.054 | 18.1 | 0.032 | 0.032 | 0.063 | .624 |
| | | DML-True | 0.011 | 3.7 | 0.033 | 0.032 | 0.035 | .932 |

(b) Latent confounder correlated with $Z_1$ ($N$ = 1,000)

| Form | ω | ρ | Estimator | Bias | RB (%) | SD | Mean SE | RMSE | Coverage |
|---|---|---|---|---|---|---|---|---|---|
| Quadratic | .63 | .6 | SEM-L | 0.153 | 50.9 | 0.037 | 0.038 | 0.157 | .014 |
| | | | SEM-Q | 0.001 | 0.4 | 0.037 | 0.038 | 0.037 | .954 |
| | | | DML-Items | 0.156 | 51.8 | 0.035 | 0.034 | 0.159 | .000 |
| | | | DML-Mean | 0.152 | 50.7 | 0.035 | 0.034 | 0.156 | .002 |
| | | | SEM-Select | 0.010 | 3.2 | 0.053 | 0.038 | 0.054 | .910 |
| Quadratic | .63 | .8 | SEM-L | 0.153 | 50.9 | 0.037 | 0.038 | 0.157 | .034 |
| | | | SEM-Q | 0.001 | 0.4 | 0.036 | 0.037 | 0.036 | .956 |
| | | | DML-Items | 0.124 | 41.5 | 0.035 | 0.033 | 0.129 | .046 |
| | | | DML-Mean | 0.122 | 40.6 | 0.035 | 0.033 | 0.127 | .040 |
| | | | SEM-Select | 0.014 | 4.5 | 0.057 | 0.037 | 0.058 | .890 |
| Quadratic | .93 | .6 | SEM-L | 0.154 | 51.3 | 0.034 | 0.034 | 0.158 | .000 |
| | | | SEM-Q | 0.002 | 0.6 | 0.032 | 0.033 | 0.032 | .956 |
| | | | DML-Items | 0.052 | 17.4 | 0.034 | 0.033 | 0.062 | .640 |
| | | | DML-Mean | 0.053 | 17.6 | 0.034 | 0.033 | 0.063 | .642 |
| | | | SEM-Select | 0.022 | 7.5 | 0.064 | 0.034 | 0.068 | .846 |
| Quadratic | .93 | .8 | SEM-L | 0.154 | 51.5 | 0.034 | 0.034 | 0.158 | .000 |
| | | | SEM-Q | 0.002 | 0.7 | 0.032 | 0.033 | 0.032 | .954 |
| | | | DML-Items | 0.049 | 16.3 | 0.033 | 0.033 | 0.059 | .668 |
| | | | DML-Mean | 0.049 | 16.3 | 0.034 | 0.033 | 0.060 | .670 |
| | | | SEM-Select | 0.023 | 7.5 | 0.065 | 0.034 | 0.068 | .846 |
| Saturating | .63 | .6 | SEM-L | 0.153 | 51.1 | 0.033 | 0.035 | 0.157 | .008 |
| | | | SEM-Q | 0.049 | 16.4 | 0.037 | 0.038 | 0.062 | .758 |
| | | | DML-Items | 0.169 | 56.4 | 0.033 | 0.032 | 0.172 | .000 |
| | | | DML-Mean | 0.164 | 54.6 | 0.033 | 0.032 | 0.167 | .000 |
| | | | SEM-Select | 0.049 | 16.4 | 0.037 | 0.038 | 0.062 | .758 |
| Saturating | .63 | .8 | SEM-L | 0.152 | 50.8 | 0.034 | 0.035 | 0.156 | .008 |
| | | | SEM-Q | 0.048 | 15.9 | 0.037 | 0.037 | 0.060 | .740 |

| Form | ω | ρ | Estimator | Bias | RB (%) | SD | Mean SE | RMSE | Coverage |
|---|---|---|---|---|---|---|---|---|---|
| | | | DML-Items | 0.137 | 45.8 | 0.034 | 0.032 | 0.141 | .004 |
| | | | DML-Mean | 0.134 | 44.7 | 0.033 | 0.032 | 0.138 | .008 |
| | | | SEM-Select | 0.048 | 15.9 | 0.037 | 0.037 | 0.060 | .740 |
| Saturating | .93 | .6 | SEM-L | 0.155 | 51.6 | 0.031 | 0.031 | 0.158 | .000 |
| | | | SEM-Q | 0.051 | 16.9 | 0.032 | 0.033 | 0.060 | .670 |
| | | | DML-Items | 0.062 | 20.6 | 0.033 | 0.032 | 0.070 | .502 |
| | | | DML-Mean | 0.058 | 19.2 | 0.033 | 0.032 | 0.066 | .564 |
| | | | SEM-Select | 0.051 | 16.9 | 0.032 | 0.033 | 0.060 | .670 |
| Saturating | .93 | .8 | SEM-L | 0.154 | 51.4 | 0.031 | 0.031 | 0.157 | .000 |
| | | | SEM-Q | 0.051 | 16.9 | 0.033 | 0.033 | 0.060 | .644 |
| | | | DML-Items | 0.058 | 19.4 | 0.034 | 0.032 | 0.067 | .566 |
| | | | DML-Mean | 0.055 | 18.4 | 0.033 | 0.032 | 0.064 | .592 |
| | | | SEM-Select | 0.051 | 16.9 | 0.033 | 0.033 | 0.060 | .644 |

(c) Proxy inputs and SEM estimators under parallel and unequal loadings (*N* = 1,000)

| Form | ω | Loadings | Estimator | Bias | RB (%) | SD | Mean SE | RMSE | Coverage | Converged (%) |
|---|---|---|---|---|---|---|---|---|---|---|
| Quadratic | .63 | Unequal | SEM-L | 0.153 | 51.0 | 0.035 | 0.037 | 0.157 | .010 | 100.0 |
| | | | SEM-Q | 0.001 | 0.5 | 0.037 | 0.037 | 0.037 | .950 | 100.0 |
| | | | DML-Items | 0.129 | 42.9 | 0.034 | 0.033 | 0.133 | .026 | 100.0 |
| | | | DML-Mean | 0.172 | 57.3 | 0.034 | 0.034 | 0.175 | .000 | 100.0 |
| | | | DML-Opt | 0.126 | 42.1 | 0.033 | 0.033 | 0.131 | .030 | 100.0 |
| | | | DML-FS | 0.130 | 43.4 | 0.034 | 0.033 | 0.135 | .028 | 92.8 |
| | | | SEM-Select | 0.013 | 4.4 | 0.054 | 0.038 | 0.056 | .890 | 100.0 |
| Quadratic | .63 | Parallel | DML-Opt | 0.171 | 57.1 | 0.035 | 0.034 | 0.175 | .002 | 100.0 |
| | | | DML-FS | 0.173 | 57.6 | 0.035 | 0.034 | 0.176 | .002 | 100.0 |
| Quadratic | .84 | Unequal | SEM-L | 0.153 | 51.1 | 0.033 | 0.035 | 0.157 | .004 | 100.0 |
| | | | SEM-Q | 0.002 | 0.5 | 0.033 | 0.034 | 0.033 | .954 | 100.0 |
| | | | DML-Items | 0.070 | 23.3 | 0.034 | 0.033 | 0.078 | .446 | 100.0 |
| | | | DML-Mean | 0.094 | 31.3 | 0.032 | 0.033 | 0.099 | .188 | 100.0 |
| | | | DML-Opt | 0.069 | 23.0 | 0.033 | 0.033 | 0.077 | .458 | 100.0 |
| | | | DML-FS | 0.070 | 23.2 | 0.033 | 0.033 | 0.077 | .444 | 100.0 |
| | | | SEM-Select | 0.020 | 6.8 | 0.060 | 0.034 | 0.063 | .854 | 100.0 |
| Quadratic | .84 | Parallel | DML-Opt | 0.094 | 31.4 | 0.033 | 0.033 | 0.100 | .200 | 100.0 |
| | | | DML-FS | 0.094 | 31.5 | 0.033 | 0.033 | 0.100 | .208 | 100.0 |

| Form | ω | Loadings | Estimator | Bias | RB (%) | SD | Mean SE | RMSE | Coverage | Converged (%) |
|---|---|---|---|---|---|---|---|---|---|---|
| Saturating | .63 | Unequal | SEM-L | 0.154 | 51.2 | 0.033 | 0.034 | 0.157 | .004 | 100.0 |
| | | | SEM-Q | 0.051 | 17.1 | 0.037 | 0.037 | 0.063 | .696 | 100.0 |
| | | | DML-Items | 0.136 | 45.4 | 0.033 | 0.032 | 0.140 | .010 | 100.0 |
| | | | DML-Mean | 0.180 | 60.2 | 0.033 | 0.032 | 0.184 | .000 | 100.0 |
| | | | DML-Opt | 0.137 | 45.7 | 0.033 | 0.032 | 0.141 | .010 | 100.0 |
| | | | DML-FS | 0.141 | 47.0 | 0.034 | 0.032 | 0.145 | .009 | 92.8 |
| | | | SEM-Select | 0.051 | 17.1 | 0.037 | 0.037 | 0.063 | .696 | 100.0 |
| Saturating | .63 | Parallel | DML-Opt | 0.181 | 60.4 | 0.034 | 0.032 | 0.184 | .000 | 100.0 |
| | | | DML-FS | 0.182 | 60.8 | 0.034 | 0.032 | 0.185 | .000 | 100.0 |
| Saturating | .84 | Unequal | SEM-L | 0.154 | 51.4 | 0.031 | 0.032 | 0.157 | .002 | 100.0 |
| | | | SEM-Q | 0.050 | 16.8 | 0.033 | 0.034 | 0.060 | .676 | 100.0 |
| | | | DML-Items | 0.079 | 26.3 | 0.033 | 0.032 | 0.085 | .312 | 100.0 |
| | | | DML-Mean | 0.103 | 34.2 | 0.032 | 0.032 | 0.108 | .114 | 100.0 |
| | | | DML-Opt | 0.076 | 25.3 | 0.033 | 0.032 | 0.083 | .350 | 100.0 |
| | | | DML-FS | 0.077 | 25.6 | 0.033 | 0.032 | 0.084 | .332 | 100.0 |
| | | | SEM-Select | 0.050 | 16.8 | 0.033 | 0.034 | 0.060 | .676 | 100.0 |
| Saturating | .84 | Parallel | DML-Opt | 0.103 | 34.5 | 0.033 | 0.032 | 0.108 | .120 | 100.0 |
| | | | DML-FS | 0.104 | 34.7 | 0.033 | 0.032 | 0.109 | .110 | 100.0 |

(d) Cubic form, in which a quadratic term recovers none of the nonlinear variance ($N = 1{,}000$)

| Form | ω | Estimator | Bias | RB (%) | SD | Mean SE | RMSE | Coverage |
|---|---|---|---|---|---|---|---|---|
| Cubic | .63 | SEM-L | 0.150 | 50.1 | 0.051 | 0.046 | 0.159 | .070 |
| | | SEM-Q | 0.128 | 42.8 | 0.045 | 0.045 | 0.136 | .162 |
| | | DML-Items | 0.200 | 66.6 | 0.049 | 0.046 | 0.206 | .002 |
| | | DML-Mean | 0.198 | 66.1 | 0.051 | 0.045 | 0.204 | .002 |
| | | SEM-Select | 0.139 | 46.2 | 0.050 | 0.046 | 0.147 | .118 |
| Cubic | .93 | SEM-L | 0.152 | 50.6 | 0.049 | 0.044 | 0.159 | .048 |
| | | SEM-Q | 0.145 | 48.3 | 0.045 | 0.040 | 0.152 | .044 |
| | | DML-Items | 0.090 | 30.0 | 0.045 | 0.041 | 0.101 | .410 |
| | | DML-Mean | 0.096 | 32.1 | 0.047 | 0.043 | 0.107 | .356 |
| | | SEM-Select | 0.148 | 49.3 | 0.046 | 0.042 | 0.155 | .048 |

*Note.* See Table C1a for column definitions; Converged = converged with a proper solution, as a percentage of attempted fits. In panel (c) the parallel rows show DML-Opt and DML-FS only: the other estimators of these conditions are in Table C1a, and DML-Opt gives the same estimates as DML-Mean there because equal weights make the optimal score a multiple of the mean and the tree-based learner is invariant to rescaling its input. In panel (d) the population bias of SEM-

L and of Reg-Q with observed $U$ is 0.153 at every reliability; SEM-Q is misspecified and its bias is reported from the simulation only.

**Figure C1**

*Root Mean Squared Error in the Core Conditions*

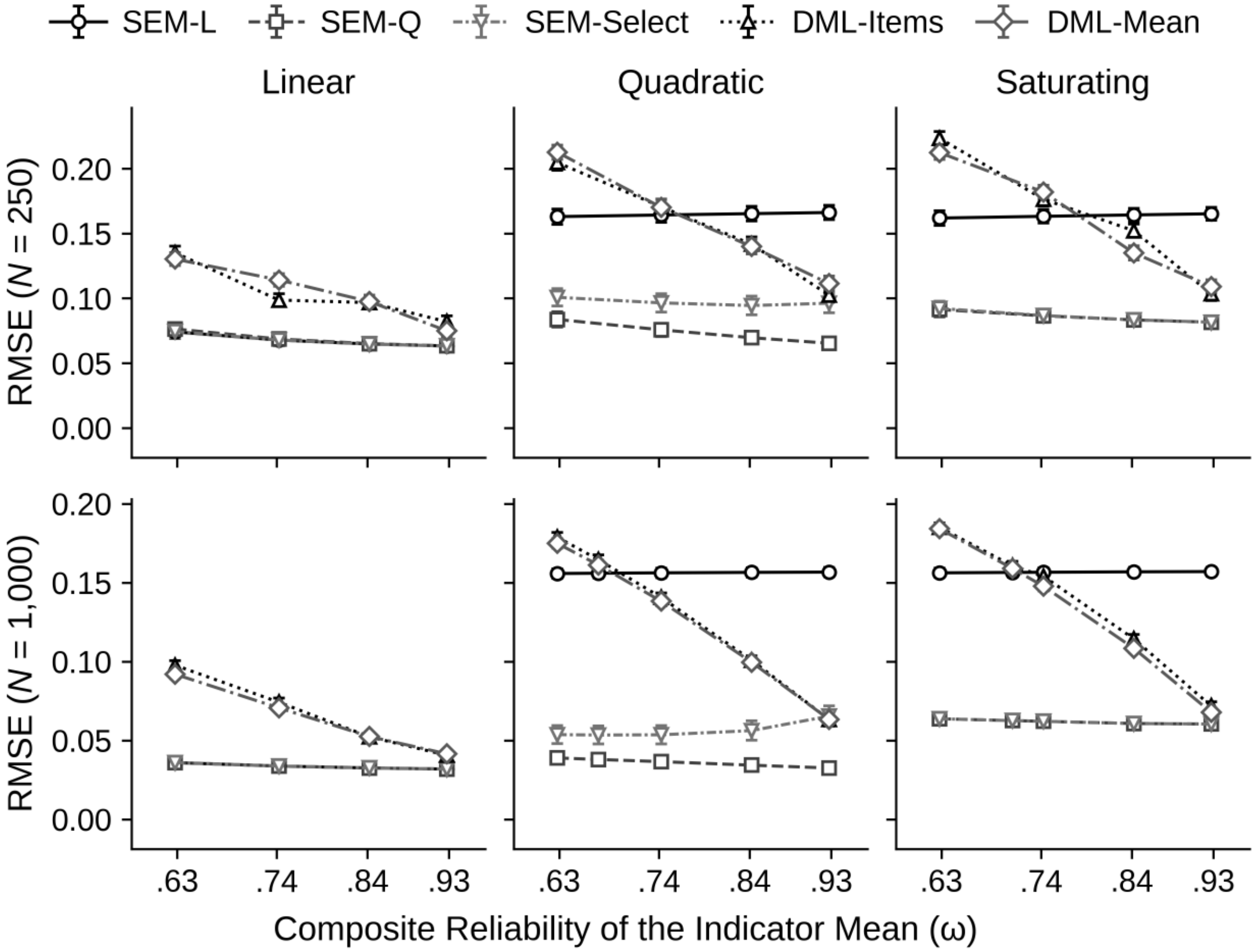

*Note.* Columns: form of confounding; rows: sample size. Points are Monte Carlo means; error bars are 1.96 Monte Carlo standard errors. RMSE equals the empirical standard deviation where an estimator is unbiased (0.032–0.039 at $N$ = 1,000; 0.063–0.083 at $N$ = 250) and is dominated by the bias elsewhere.

## Supplement D. Preregistration

### D.1 Original Registration (24 February 2026)

The study was registered on the Open Science Framework (OSF) on 24 February 2026 (embargoed until publication; anonymized view-only link in the main text). The registration specified three hypotheses, on the nonlinear SEM, on DML under adequate versus poor reliability, and on coverage under misspecification; a 2 (functional form) × 3 (indicator reliability 1.0, .81, .49) × 2 ($N$ = 250, 1,000) design with 500 valid replications and seed 20260224; a nonlinear form with a quadratic and a latent-by-observed interaction term; SEM-L, SEM-NL by LMS with 7 integration points, and DML with three inputs, tuned inside each replication; and resampling of datasets with failed fits.

**D.2 Amendment (10 October 2026)**

The amendment was deposited on OSF as a single document, next to the unchanged original registration, before the pilot and main runs; when it was deposited, no estimator had been fitted to a dataset of the amended design and no result had been inspected. It states that peer review raised four issues (the latent confounder was independent of every covariate; the indicators had equal loadings; the reliability at which the estimators changed order was interpolated from two conditions; the nonlinear SEM was specified with the exact generating terms), replaces the hypotheses by the three research questions, and fixes in advance how failed fits and machine-learning tuning are handled. Its protocol section specifies the data-generating model, the 46 conditions, the 500 replications and seed, the estimators, the tuning protocol, the usable-fit rule, the performance measures, the research questions, and the software exactly as the Method and Supplement B describe them (in the notation of the deposited document, $c$ denotes the nonlinear coefficients $\kappa_T$, $\kappa_Y$; $e_T$, $e_Y$ the disturbances $\zeta_T$, $\zeta_Y$; the "odd" block is the cubic block and the "congeneric" block the unequal-loadings block). Its list of changes and its rules are reproduced verbatim (the bracketed words are added).

*Changes from the original registration.*

- Nonlinear confounding: quadratic plus latent-by-observed interaction → the four forms above [linear, quadratic, saturating, cubic]; the interaction was dropped.
- Measurement: indicator reliabilities 1.0, .81, .49 → composite reliabilities .63–.93, with crossover, congeneric and observed-confounder conditions.
- Latent confounder and covariates: independent → correlated block added; *U* regressed on *Z* in every SEM.
- Design: 12 → 46 conditions; structural coefficients, software and seed (20260224 → 20261009) fixed.
- Replications: 500 valid datasets with resampling of failed fits → 500 generated datasets, failures counted.
- Latent models: SEM-NL with interaction and quadratic terms, 7 integration points → SEM-Q with the quadratic term only, 30 points; SEM-Select added.
- DML: 3-fold cross-fitting and a 15-candidate search inside each replication → 5-fold cross-fitting and tuning fixed in advance; PCA fallback removed; DML-Opt and DML-True added.
- Hypotheses H1–H3 → research questions 1–3.

*Rules fixed in advance.*

1. The conditions, replications, seed, estimators, tuning protocol and performance measures above are final; the replication count will not change after any main-run result has been inspected.
2. Hyperparameters are selected before the pilot and main runs on tuning datasets from a separate seed stream, never from datasets that enter the results.
3. A pilot of seven conditions with two replications each verifies the software chain; its results are not reported. After the pilot, the only setting that may change is a time limit shown to be binding for fits that are still iterating normally; the new limit and the reason are recorded before the main run.

4. Failed fits are reported by status; no fit is rerun to obtain a different result. Software or input errors are retried automatically and, if they persist, repaired and rerun explicitly, with all attempts retained.
5. Differences between estimators smaller than two paired Monte Carlo standard errors are not interpreted; the ordering of DML and SEM-L is called "predicted by the crossover reliability" only when the paired difference has the predicted sign and exceeds that margin.